\documentclass[iop]{emulateapj}

\usepackage{natbib} 
\newcommand{\unit}[1]{\mathrm{#1}}

\newcommand{\wpp}{w_{\mathrm{p}}}
\newcommand{\wprp}{w_{\mathrm{p}}(r_{\mathrm{p}})}
\newcommand{\rp}{(r_{\mathrm{p}})}

\newcommand{\Mpc}{\unit{Mpc}} 

\newcommand{\Gyr}{\unit{Gyr}}

\newcommand{\kms}{\unit{km \ s^{-1}}}

\newcommand{\hmpc}{h^{-1}\mathrm{Mpc}}
\newcommand{\hkpc}{h^{-1}\mathrm{kpc}}
\newcommand{\hMsun}{h^{-1}M_{\odot}}

\newcommand{\Mmin}{M_\mathrm{min}}
 
\newcommand{\Mstar}{M^{\ast}}
\newcommand{\Lstar}{L^{\ast}}
\newcommand{\MDM}{M^\mathrm{dm}}

\newcommand{\Vmax}{V_\mathrm{max}}
\newcommand{\Vmaxacc}{V_\mathrm{max}^\mathrm{acc}}

\newcommand{\xir}{\xi(r)}

\newcommand{\Msun}{M_{\odot}}
 
\newcommand{\rinfl}{r_{\mathrm{infl}}} 

\newcommand{\rprime}{r'}

\newcommand{\fihlbhg}{f_\mathrm{IHL+BHG}}
\newcommand{\fihl}{f_\mathrm{IHL}}
\newcommand{\pimax}{\pi_\mathrm{max}}
\newcommand{\fDMloss}{f_\mathrm{DM \ loss}}

\usepackage{epsfig} 
\usepackage{graphicx}  
\usepackage{rotating}

\begin{document}

\title{CONSTRAINING SATELLITE GALAXY STELLAR MASS LOSS AND PREDICTING INTRAHALO LIGHT I:  
FRAMEWORK AND RESULTS AT LOW REDSHIFT}

\author{Douglas~F.~Watson and Andreas~A.~Berlind\altaffilmark{1},}
\affiliation{Department of Physics and Astronomy, Vanderbilt University, Nashville, TN 37235}

\author{Andrew~R.~Zentner} \affiliation{
Department of Physics and Astronomy \& 
Pittsburgh Particle physics, Astrophysics, and Cosmology Center (PITT PACC), 
The University of Pittsburgh, Pittsburgh, PA 15260
}

\begin{abstract}
We introduce a new technique that uses \emph{galaxy clustering} to
constrain how satellite galaxies lose stellar mass and contribute to
the diffuse ``intrahalo light'' (IHL).  We implement two models that
relate satellite galaxy stellar mass loss to the detailed knowledge of
subhalo dark matter mass loss.  Model~1 assumes that the 
fractional stellar mass loss of a galaxy, from the time of merging 
into a larger halo until the final redshift, is proportional to the 
fractional amount of dark matter mass loss of the subhalo it lives in.  
Model~2 accounts for a delay in the time that stellar mass is lost due 
to the fact that the galaxy resides deep in the potential well of the 
subhalo and the subhalo may experience dark matter mass loss for some 
time before the galaxy is affected.  We use these models to predict
the stellar masses of a population of galaxies and we use 
\emph{abundance matching} to predict the clustering of several $r$-band 
luminosity threshold samples from the Sloan Digital Sky Survey.  Abundance matching 
assuming no stellar mass loss (akin to abundance matching at the time 
of subhalo infall) over-estimates the correlation function on 
small scales ($\lesssim 1\Mpc$), while allowing too much stellar mass 
loss leads to an under-estimate of small-scale clustering.  
For each luminosity threshold sample, we are thus able to 
constrain the amount of stellar mass loss
required to match the observed clustering.  We find that satellite
galaxy stellar mass loss is strongly luminosity dependent, with less
luminous satellite galaxies experiencing substantially more efficient 
stellar mass loss than luminous satellites.  
With constrained stellar mass loss models, we can infer the amount of 
stellar mass that is deposited into the IHL.  We find that both of
our model predictions for the mean amount of IHL as a function of halo 
mass are consistent with current observational measurements.  
However, our two models predict a different amount of scatter in the IHL 
from halo to halo, with Model~2 being favored by observations.  
This demonstrates that a comparison to IHL measurements provides independent 
verification of our stellar mass loss models, as well as additional
constraining power.

\end{abstract}


\altaffiltext{1}{Alfred P. Sloan Fellow}


\section{INTRODUCTION}\label{intro}


In the concordance $\Lambda$CDM cosmology, galaxies,
galaxy groups, and galaxy clusters form hierarchically.  
High-density regions condense and virialize, 
forming bound structures known as halos.  
Halos grow through the continual accretion of 
smaller objects.  These accreted objects 
may survive within the virialized region of the primary halo 
as smaller, self-bound, orbiting dark matter clumps dubbed 
``subhalos'' or ``substructure'' \citep{Ghigna00,klypin99a,diemand04,kravtsov04a}.
Halos of sufficient mass are the natural sites of galaxy formation, with 
baryons cooling and condensing towards potential well minima 
\citep{whiterees78,blumenthal86}.  When a halo is accreted by a larger 
halo, thus becoming a subhalo, the galaxy within it becomes a ``satellite'' 
galaxy within a group or cluster.   Understanding the detailed relationship 
between (satellite) galaxies and (sub) halos is a long-standing focus of galaxy 
formation theory.


In the hierarchical paradigm, these smaller objects, upon merging,
become victims  of intense tidal fields and interactions within the
larger systems in which they reside.   The dark matter mass associated
with a subhalo may be rapidly stripped upon  infall.  This stripping
acts on the periphery of the subhalo first,  suggesting that the
luminous galaxy, residing in the center of the subhalo,  may be
relatively unharmed.  After enough time has elapsed, stripping of
stars may begin  to occur as well.   These liberated stars that are
ripped from galaxies are the  likely source of ``intrahalo light''
\citep[IHL:
e.g.,][]{gallagher_ostriker72,merritt83,byrd_valtonen90,gnedin03,murante04,
lin_mohr04,willman04,sommerlarsen06,conroy07,skibba07,purcell_etal07,purcell_etal08,yang09b,
rudick09,rudick11}.   This has been studied in great detail \emph{at
the scale of individual galaxies}, often categorized as
"stellar halos" or "tidal streams"
\citep{morrison93,sackett94,wetterer_mcgraw96,morrison97,weil97,chiba_beers00,
ivezic00,lequeux98,abe99,morrison00,yanny00,siegel02,irwin05,zibetti_ferguson04,
guhathakurta05,chapman06,kilarai06,McConnachie06,hood07,mackey10,bailin11},
\emph{galaxy groups}
\citep{feldmeier01,castrorodriguez03,white_P03,darocha05,aguerri06,feldmeier06,darocha08},
and \emph{galaxy clusters} \citep[where it is known as the
intra-cluster light, or ICL, e.g.,
][]{gallagher_ostriker72,merritt83,melnick77,thuan_kormendy77,byrd_valtonen90,
uson91,bernstein95,calc_rold00,gnedin03,murante04,lin_mohr04,willman04,mihos05,
zibetti05,krick06,sommerlarsen06,conroy07,seigar07,gonzalez07,pierini08,rudick09,
rudick11,toledo11,romanowsky12}.

In this paper, we seek to understand the liberation of stars from
satellite galaxies by connecting stellar mass loss\footnote{By ``stellar
mass loss'', we refer to stars being stripped from a galaxy and not
gas lost from stars via winds.} to subhalo dark matter mass loss using 
\emph{galaxy clustering} and \emph{intrahalo light} observations.  
We employ the model for halo substructure 
introduced in \citet[][hereafter Z05]{zentner05} in order to  
constrain this relationship.  We compare our model predictions to observations 
of the IHL over a large range of host halo mass scales.  The aim of this 
paper (Paper I) is to introduce our modeling framework and its predictive 
power.  In a forthcoming paper (Paper II) we will extend our analysis to high
redshift in order to study the assembly of the IHL across cosmic time.


The paper is laid out as follows.  In \S~\ref{sec:motivation} we
discuss the motivation for this study. In \S~\ref{sec:dynamical_model}
we review the Z05 model for cold dark matter (CDM) substructure.  
In \S~\ref{sec:stripping_models} we describe our models that connect
stellar mass loss to dark matter mass loss and in 
\S~\ref{sec:compare_clustering} we demonstrate how we constrain these models
using galaxy clustering.  In \S~\ref{sec:lum_dependence} we
show the luminosity dependence of satellite galaxy stellar mass loss.
In \S~\ref{sec:predict_IHL} we use our models to make IHL predictions, 
and compare to observations at low redshift.  Finally, in 
\S~\ref{sec:conclusion} we give a summary of our results and discuss 
directions for future work.
%


\section{MOTIVATION}\label{sec:motivation}


A simple, yet remarkably powerful technique for connecting dark matter
halo mass to either stellar mass or luminosity has emerged in recent 
years. By assuming a monotonic relation between halo mass (or
maximum circular velocity $\Vmax = \mathrm{max}[\sqrt{GM(<r)/r}]$) 
and luminosity $L$ (or stellar
mass) one can ``abundance match'' to make the correspondence between
dark matter (sub)halos and an observable galaxy property
\citep[e.g.,][]{kravtsov04a,vale_ostriker04,tasitsiomi_etal04,vale_ostriker06,
conroy06,conroy_wechsler09,moster10,behroozi10,guo10,simha10}.
For example, this can be done by matching the observed number density
of galaxies, $n_{g}$, above some luminosity to the number density of
halos and subhalos, $n_{h}$, above a certain $\Vmax$,
\begin{equation}
n_{g}(>L) = n_{h}(>V_\mathrm{max}) .
\end{equation}
This yields an implicit relationship between $L$ and $\Vmax$ that
preserves the observed luminosity function of galaxies.

\citet{conroy06} used this method to assign luminosities to halos and
subhalos in a cosmological N-body simulation at several redshifts.
They predicted the luminosity-dependent clustering of galaxies
and found that the two-point correlation function (2PCF) of halos and 
subhalos matched that of galaxies for a wide range of luminosities
and redshifts.  
The authors made the physically-motivated choice
that $\Vmax$ should be the maximum circular velocity of subhalos 
\emph{at the time of accretion}, $\Vmaxacc$, instead of at the
time of observation \citep[see also][]{nagai05,vale_ostriker06,berrier06}.

The reasoning behind this choice is as follows.  Upon merging into a
larger  host, the $\Vmax$ of a subhalo will decrease due to mass loss
\citep{hayashi03,kravtsov04a}.  Dark matter on the periphery  of the
subhalo will be lost first, because it is less bound to the subhalo.
On the other hand, the stellar mass of the satellite galaxy  is
concentrated at the center of the subhalo and thus more tightly bound.
Tidal stripping can significantly alter the surrounding subhalo, but
possibly  leave the galaxy largely unperturbed for some period of time
(using N-body simulations, \citealt{P08b} found that halos need to
lose $\sim90\%$ of their original mass before tides begin to affect the
kinematics of stars).  Consequently, while $\Vmax$ of the subhalo
decreases, the stellar mass  of the galaxy may remain unchanged for
long periods of time.  It follows that galaxy observables such as
luminosity or stellar mass should correlate  with $\Vmaxacc$ instead
of the final $\Vmax$ \citep{nagai05}.  \citet{conroy06}  lent
empirical support to this picture by showing that the choice of
$\Vmaxacc$ was essential in order to achieve agreement with the
observed clustering of galaxies.

It is well known that galaxy groups and clusters are replete with
diffuse stellar material.  This material is widely thought to be the 
remains of disrupted satellites (see \S~\ref{intro}). 
However, the choice to associate galaxy luminosity with $\Vmax$  
at the time of satellite accretion is tantamount to assuming that 
no stars become unbound from satellite galaxies. In this case, 
why can using $\Vmaxacc$ accurately reproduce the observed clustering?

There is a crucial subtlety to simply using $\Vmaxacc$ for abundance
matching.   When matching galaxy  and subhalo number densities with
the $L - \Vmax$ relation by choosing  $\Vmaxacc$, there is an inherent
second threshold in the final $\Vmax$ of subhalos. This is a threshold
below which objects are ``operationally removed from consideration''
due to the limited resolution of the simulation that is used to
perform the calculation.  In the study of \citet{conroy06}, this
threshold was $\sim 80\kms$ -- the completeness limit of halos in
their simulation.  Therefore, an accreted  subhalo will be
artificially destroyed if it becomes sufficiently small such that its
structure is not well resolved within the simulation.  These so-called
``orphans'' are neglected and they unwittingly act to model the
stripping of stars.  The galaxies that would have been associated with
these halos, had they not become unresolved, are effectively removed
from the final galaxy sample, just as if they were disrupted.  This
elimination of subhalos directly affects the 2PCF, which is very
sensitive to subhalo abundance \citep[see][]{watson_powerlaw11}.
Including these \emph{orphans} should result in a boost of the
small-scale correlation function, implying that pure abundance
matching  using $\Vmaxacc$ may not accurately reproduce the observed
clustering on small scales
\citep[see][]{kitzbichler08,moster10,wetzel_white10}.  In fact, we
have performed a test in which we selected a $\Vmaxacc > 210 \kms$
(corresponding to $\sim \Lstar$ galaxies and brighter) threshold with
two, secondary final $\Vmax$ thresholds, $\Vmax > 20 \kms$ and  $\Vmax
> 80 \kms$, to mimic this resolution limit effect.  Increasing this
threshold from $20$ to $80 \kms$ resulted in a $\sim 20\%$ decrease in
the 2PCF\footnote{We explain the tools we use to calculate correlation
functions in \S~\ref{sec:compare_clustering}.} at scales less than
$1\Mpc$.  We speculate that below this resolution limit is where  much
of the stripped stellar material may originate.
%

The above discussions bring us to the motivation of this work. Including
stellar mass loss in the abundance matching technique is necessary to
describe the observed clustering of galaxies in a manner that does not
include implicit selections.   Observed galaxy clustering therefore
has the potential to constrain  the typical amount of stellar mass
loss from a subhalo.   In this work, we develop models that relate
satellite galaxy stellar mass loss to subhalo dark matter mass loss.
We use these models, together with the abundance matching technique,
to predict the observed  2PCF (see \S~\ref{sec:stripping_models} and
\S~\ref{sec:compare_clustering}  for details).  Comparing to
measurements allows us to constrain our stellar mass loss models.
This requires a detailed understanding of the evolution  of subhalos
within hosts.  We use the Z05 analytic model for halo
substructure, which is not subject to any intrinsic resolution
effects.   The model is capable of tracking surviving subhalos down to
$\Vmax \ll 80\ \kms$, so predictions are not affected by the
``orphan'' population of galaxies.  Our models yield stellar mass loss
histories from satellite galaxies, so we can make predictions for the
amount of intrahalo light at varying scales and compare to
observations.  This provides an independent check of our stellar mass
loss constraints.  This modeling can also be done at varying redshifts
in order to study the evolution of stellar mass loss and the assembly
of IHL over time. Ultimately, this investigation could enable a single 
model to track the stellar-to-halo mass relation as a function of time 
such that abundance matching at multiple redshifts would be unnecessary.

\section{THE SUBHALO EVOLUTION MODEL}\label{sec:dynamical_model}


We construct models in which the stellar mass loss of a satellite
galaxy is linked to the dark matter mass loss of its subhalo.
Consequently, we  require a detailed understanding of the evolution of
subhalos within hosts.   For this we use the Z05 model, which is based
on \citet{zentner03} and is similar to the independent models of
\citet{TB04a,TB05b,TB05c} and \citet{PenarrubiaBenson05}, and shares
many features with other approximate treatments of subhalo populations
\citep{OguriLee04,vdBosch05,faltenbacher_matthews05,purcell_etal07,
giocoli_etal08,giocoli_etal09,gan10,yang11a}.  The Z05 model produces subhalo 
mass functions, occupation statistics, and radial distributions within 
hosts that are in good agreement with a number of high-resolution $N$-body 
simulations \citep[Z05, and the recent comparison in][]{koushiappas_etal10}.


The analytic model proceeds as follows.  For a host halo at a given
redshift $z$, and mass $M$, we generate a halo merger tree
using the mass-conserving implementation of the excursion set
formalism (\citealt{bond91,LaceyCole93,LaceyCole94}; see 
\citealt{zentner07} for a review) developed by \citet{somerville99}.   
This provides the entire history of all halos that merged to form the 
final, host halo.  We assign all halos \citet[][hereafter NFW]{nfw97} 
density profiles with concentrations determined by their merger 
histories according to \citet{wechsler02}.  When a halo merges into a 
larger halo, it becomes a subhalo and is assigned initial orbital 
parameters drawn from distributions measured in N-body simulations 
(Z05; also see \citealt{benson05} for similar formalisms).  We then 
integrate each subhalo orbit within the gravitational field of the host, 
subjecting the subhalos to orbital decay via 
dynamical friction and mass loss through tides and interactions.  
We estimate dynamical friction with an updated form of the 
\citet{chandrasekhar43} approximation \citep{hashimoto_etal03,zentner03},
and allow for dark matter mass loss beyond the tidal radius on a timescale
comparable to the local dynamical time.  Finally, we account for internal 
heating so that scaling relations describing the internal structures of 
subhalos are obeyed \citep{hayashi03,kazantzidis_etal04,kravtsov04b}.  
We refer the reader to Z05 for specific details of these ingredients.

The lack of a resolution limit in the Z05 model allows us to track
subhalos regardless of how much mass they have lost.  Therefore, we do
not lose subhalos due to mass loss.  However, we do consider a subhalo
to be effectively ``destroyed'' if its orbital apocenter sinks to less
than $r_{\rm apo} < 5$~kpc distance from the center of its host halo.
This choice is physically motivated because the galaxy within such a
subhalo would likely have merged with the central galaxy, or at least
be observationally indistinguishable from it.  This criterion thus
models  the cannibalism of satellite galaxies by central galaxies.  We
show in  \S~\ref{sec:predict_IHL} that inclusion of these cannibalized
subhalos  in our predictions for intrahalo light is negligible, so
this criterion  actually has little effect on our modeling results.
In practice, subhalos  lose the vast majority of their mass prior to
achieving such small  apocenters, so this cut serves only to
terminate the integration of  that particular orbit.  In the end, we
amass catalogs of all  surviving subhalos in the final host halo at
the  time of ``observation'' and we know exactly how much dark matter
has been lost from each subhalo.  A halo that merges into a larger
host may  contain subhalos of its own.  These ``subs-of-subs'' of
sufficiently  high mass to host an observable galaxy are only abundant
inside very large host masses.  They are present in our model, but
rare.


\begin{figure*}[t]
\begin{center}
\includegraphics[width=1.\textwidth]{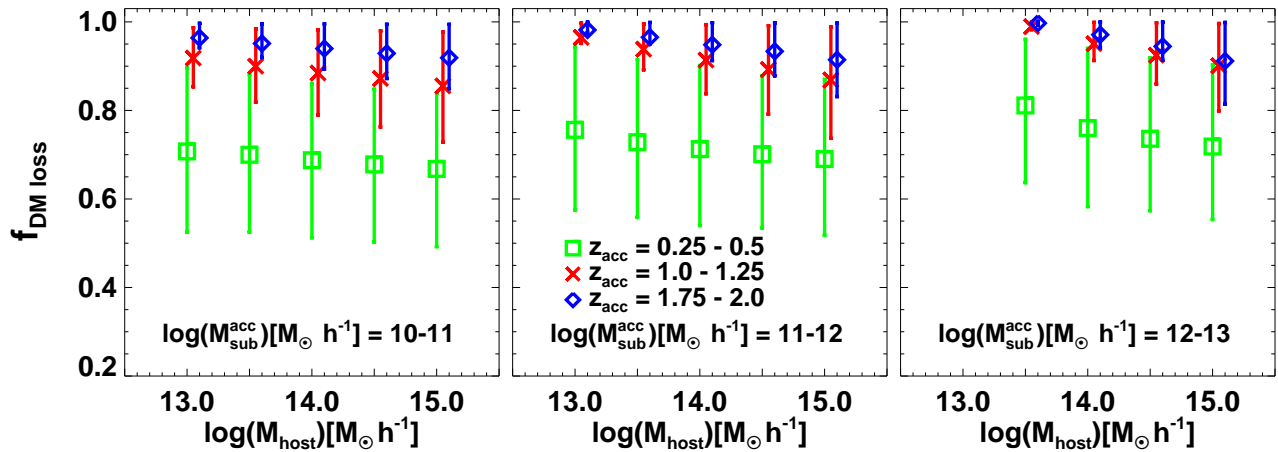}
\caption{The average fraction of dark matter lost from subhalos,
$\fDMloss = (\MDM_{\mathrm{acc}} - \MDM_{\mathrm{fin}})/\MDM_{\mathrm{acc}}$, 
as a function of host halo mass for three bins of subhalo accretion 
epoch $z_{\mathrm{acc}}$ (shown by the three point types), and three 
bins of subhalo mass (at accretion; shown by the three panels), 
according to the Z05 semi-analytic model.  Each point represents 
an average over all subhalos in a given bin of accretion epoch and 
subhalo mass, from 500 model realizations of a specific host mass, 
and errorbars show the $1\sigma$ scatter.  The points representing 
different accretion epoch bins are slightly staggered for clarity.  
Accretion time is the dominant factor that determines $\fDMloss$, 
with subhalos that have merged earlier having more time to be 
stripped of their dark matter.}
\label{fig:frac_mass_loss}
\end{center}
\end{figure*}


To properly sample the distribution of halos in the universe, we compute 
subhalo populations for a grid of host halo masses in the range $ 11 \le \log (M/\hMsun) \le 15$ 
(in steps of 0.1).  To account for statistical variation among halos and subhalos, 
we perform 500 statistical realizations of the subhalo population at each host 
mass.  The result is 500 host halos along with their subhalos at each of 
41 distinct masses, giving a total of 20,500 distinct subhalo populations.  
The model predicts the amount of dark matter lost from each 
subhalo as it orbits in the tidal field of its host halo. 
Figure~\ref{fig:frac_mass_loss} shows the fraction of subhalo mass lost 
as a function of host halo mass, in bins of accretion epoch 
$z_{\mathrm{acc}}$, and subhalo mass (at accretion).  This fraction is 
defined as $\fDMloss = (\MDM_{\mathrm{acc}} -
\MDM_{\mathrm{fin}})/\MDM_{\mathrm{acc}}$, where $\MDM_{\mathrm{acc}}$
is the subhalo mass at accretion and $\MDM_{\mathrm{fin}}$  is the
final subhalo mass.   Each panel of the figure represents a bin of
subhalo mass and the three sets of points in each panel represent 
bins of accretion epoch:  $z_{\mathrm{acc}}=0.25-0.5$ (green circles),
$z_{\mathrm{acc}}=1.0-1.25$ (red $X$ symbols), and
$z_{\mathrm{acc}}=1.75-2.0$ (blue diamonds).  Each point thus represents
an average over all subhalos in a bin of accretion epoch and subhalo 
mass, from the 500 model realizations of a specific host halo mass, and 
errorbars show the $1\sigma$ scatter.  It is clear that accretion time 
is crucial towards determining $\fDMloss$, with subhalos that 
have merged earlier having more time to be stripped of their dark matter.
Moreover, the scatter in $\fDMloss$ shrinks for earlier
accretion times.  At fixed accretion epoch, the average value of 
$\fDMloss$ is remarkably constant.  However, there is a 
slight increase as we move to lower host halo mass at fixed subhalo mass,
or as we move to higher subhalo mass at fixed host mass.  This is a
result of dynamical friction playing a stronger role for subhalos that are considerable 
in size compared to their host halos.  The larger a subhalo relative to its host, 
the more rapidly its orbit will decay and it will sink within the host potential.  
The tidal field of the host is stronger towards the host center, 
and this induces more dark matter mass loss. 

\begin{figure*}
\begin{center}
\includegraphics[width=.85\textwidth]{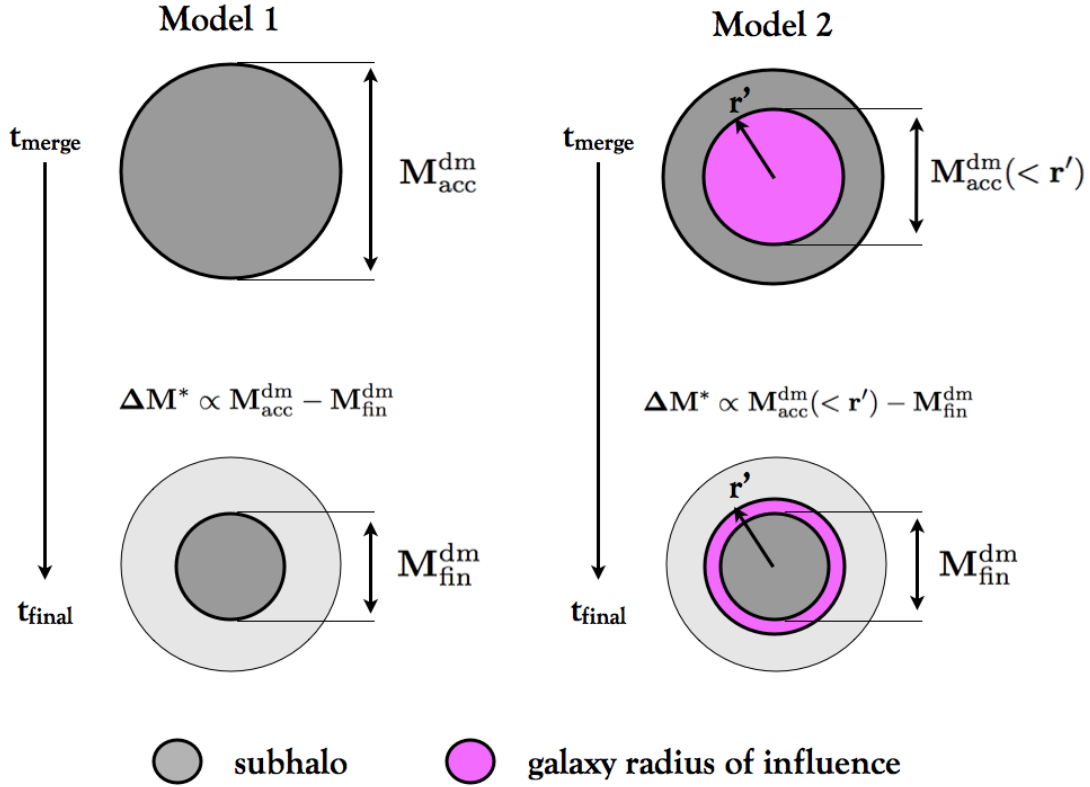}
\caption{Schematic of how satellite galaxy stellar mass loss occurs in
Model~1 and Model~2.  From the time the subhalo merges into the host
halo,  $t_\mathrm{merge}$, until the final redshift under
consideration, $t_\mathrm{final}$,  our semi-analytic subhalo model
predicts the amount of dark matter mass loss,  $\Delta \MDM \equiv
\MDM_{\mathrm{acc}}-\MDM_{\mathrm{fin}}$.  Model~1 assumes that  the
amount of stellar mass lost, $\Delta \Mstar$, is proportional to
$\Delta \MDM$.   Model~2 defines a radius, $\rprime$, such that
stellar mass is only lost if dark  matter is lost within that radius.
If this is the case, then $\Delta \Mstar$ is  only proportional to the
amount of dark matter lost inside of this radius.  See
\S~\ref{sec:stripping_models} for detailed descriptions of the models.}
\label{fig:cartoon}
\end{center}
\end{figure*}



\section{MODELS FOR SATELLITE GALAXY STELLAR MASS LOSS}\label{sec:stripping_models}


We now present the models we use to describe satellite galaxy stellar 
mass loss.  The exact way in which stars are stripped from the subhalo 
they inhabit is likely quite complicated.  However, we can make some 
simple, physically-motivated approximations aimed at capturing the 
gross, relevant behavior.  In this sense, we pursue the question 
of the evolution of stellar mass using a philosophy similar to that 
which underlies abundance matching.  We aim to make a set of minimal, 
yet effective assumptions that serve to distill the enormous amount of 
information contained in survey data.  Indeed, we aim in part to extend 
the abundance matching techniques by making the lower threshold 
for stellar mass explicit, rather than implicit.  

We consider two models in which we relate the  amount of satellite
galaxy stellar mass lost to the corresponding  amount of subhalo dark
matter lost.  Combined with the Z05 model  that makes detailed
predictions for  dark matter mass loss, these models can predict the
stellar mass loss for any given halo.  It is important to note that
the Z05 prescription for mass loss does not include the effects that
baryons can have on dark matter.  \emph{Adiabatic Contraction} (AC)
\citep{blumenthal86,ryden_gunn87,gnedin04} may increase the central
density as the gas condenses and sinks to to the center of the dark
matter potential well
\citep{diemand04,fukushige04,reed05,delpopolo09}. Conversely,
processes during halo formation, such as gravitational heating from
merger events, can counteract AC \citep[e.g.,][]{zappacosta06} and, in
fact, the central density may decrease through baryonic feedback
\citep{governato12}.  These effects can impact the survival of
satellite galaxies \citep[see][]{PenarrubiaBenson10}, however
consideration of these detailed effects in our modeling is beyond the
objective of this paper.


\subsection{Model 1}\label{model_1}


Our first model sets the fraction of stellar mass that
is lost from a galaxy to a fixed proportion of the fraction of 
dark matter that is lost from its subhalo.  The model works as follows.  
Any halo of sufficiently large dark matter mass will have some stellar mass 
associated with the galaxy it contains at the time of accretion,
$\Mstar_{\mathrm{acc}}$.  This stellar mass will be some fraction of
the mass in dark matter, $\MDM_{\mathrm{acc}}$.  After the halo merges
into a larger halo, becoming a subhalo, it orbits within the host halo
potential and loses mass.  At the time of
observation, the subhalo has a smaller mass, $\MDM_{\mathrm{fin}}$.
We relate the fraction of stellar mass that is lost during this time
to the fraction of dark matter mass that is lost through a single 
parameter $\epsilon$,
\begin{equation}
\frac{\Delta \Mstar}{\Mstar} = \epsilon \ \frac{\Delta \MDM}{\MDM},
\end{equation}
\begin{equation}
\frac{\Mstar_{\mathrm{acc}} -
\Mstar_{\mathrm{fin}}}{\Mstar_{\mathrm{acc}}} = \epsilon \
\frac{\MDM_{\mathrm{acc}} - \MDM_{\mathrm{fin}}}{\MDM_{\mathrm{acc}}},
\end{equation}
which can be re-written as,
\begin{equation}\label{eqn:main_equation}
\Mstar_{\mathrm{fin}} =\Mstar_\mathrm{acc} \left[1 -  \epsilon \times 
\Big( 
\frac{\MDM_{\mathrm{acc}} -
\MDM_{\mathrm{fin}}}{\MDM_{\mathrm{acc}}}\Big) \right] 
\end{equation}

The left-hand side of Figure~\ref{fig:cartoon} is a cartoon schematic of
how stellar mass loss occurs in Model~1.  From the time the subhalo
merges into the host halo, $t_\mathrm{merge}$, until the final
redshift under consideration, $t_\mathrm{final}$, the Z05 model (see
\S~\ref{sec:dynamical_model} for model details) predicts the amount of
subhalo dark matter mass loss $\Delta \MDM$ with the stellar mass loss
$\Delta \Mstar$ being related to the dark matter mass loss via
$\epsilon$. Therefore, we are left with a simple parametric equation
governed by a single free parameter.  For example, if $\epsilon=0.5$, 
then a subhalo that loses 50\% of its dark matter will lose 25\% of 
its stellar mass.  As we mentioned above, stellar mass loss should be 
less efficient than dark matter mass loss, so we should generally 
expect $\epsilon < 1$. 

\begin{figure}
\begin{center}
\includegraphics[width=.5\textwidth]{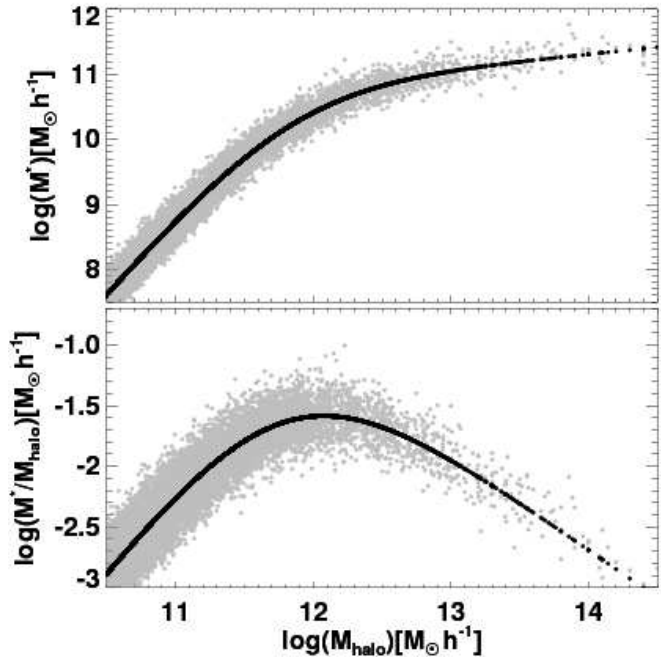}
\caption{The stellar-to-halo mass relation (SHMR) for our host halos
and subhalos with the adopted \citet{behroozi10} SHMR.  The black
curve shows the mean of the SHMR and grey points illustrate the
assumed 0.15 dex scatter by sampling 1 million halos and subhalos from
all of the model realizations.  The bottom panel is the same as the top
with the y-axis divided by halo mass to highlight the characteristic
mass ($\sim 10^{12} \hMsun$) where star formation is most efficient.
}
\label{fig:SHMR_Behroozi}
\end{center}
\end{figure}


In order to compute the final stellar mass of a galaxy via
Eqn.~\ref{eqn:main_equation}, we need to know the stellar mass of a
satellite galaxy as a function of the subhalo mass at the time of
accretion.  The stellar mass of a galaxy depends on many physical
processes, including mergers, gas cooling, star formation, feedback
from supernovae, feedback from active galactic nuclei,  making
\emph{ab initio} predictions highly non-trivial.  However, there are
several empirical methods for obtaining the stellar-to-halo mass
relation (SHMR), which have recently appeared  in the literature
\citep[e.g.,][]{yang03,wang06,
conroy_wechsler09,wang_jing10,moster10,guo10,behroozi10,neistein11}.
We assign stellar masses to halos with the SHMR of \citet[][hereafter
B10]{behroozi10}. Specifically, we employ the relation used in Eqn.~21
of B10, with the mean parameter values given by the $\mu = \kappa = 0$
model in their Table~2.  Therefore, for every halo and subhalo in our
catalog, we can use the halo mass \emph{at the time of accretion} to
assign a stellar mass to the galaxy hosted by the halo.  The top panel
of Figure~\ref{fig:SHMR_Behroozi} shows how stellar masses are related
to our host halos and subhalos with the B10 SHMR.  The black curve
shows the mean of the SHMR and grey points illustrate the assumed 0.15
dex scatter given in B10.  We show one million points randomly drawn
from the full distribution of halos and subhalos over the host
halo mass range $ 11 \le \log (M/\hMsun) \le 15$.  
%
Stellar mass rapidly increases as a function of halo mass at
low masses before turning over and becoming shallower at higher host
masses.   The bottom panel is the same as the top with the y-axis
divided by halo mass in order to highlight the characteristic mass
where this turnover occurs, $M_{\rm halo} \sim 10^{12} \hMsun$.  This
characteristic mass is the halo mass at which  star formation is most
efficient.

For most applications, SHMRs are developed by linking the halo mass
function to the galaxy stellar mass function (SMF) through abundance
matching.  This is typically done using the subhalo mass at accretion
which, by definition, presumes that no stellar stripping occurs.
However, even if stellar stripping occurs (i.e., if $\epsilon$ is
non-zero) we can still use a SHMR relation that assigns stellar mass
at the time the subhalo merges, due to the fact that the SMF is
\emph{strongly} dominated by central galaxies.  We emphasize that
while there are many SHMRs in the literature, we are not very
sensitive to the particular choice of SHMR.  As we will discuss in
\S~\ref{sec:compare_clustering}, we assign stellar masses to halos and
subhalos and then rank them in stellar mass in order to find the
stellar mass cut-off ($\Mstar_{\mathrm{fin}}$) that matches the
observed number density of a given galaxy luminosity threshold sample.
Therefore, two SHMRs that yield the same rank order for the halos and
subhalos will be  indistinguishable from each other, even though the
values of  $\Mstar_{\mathrm{fin}}$ will be different (as well as the
mass-to-light ratios).  All published SHMRs are  monotonically
increasing functions, so this behavior is general.  On the other
hand, the scatter in the SHMR is important, as this will change  the
rank order.  We use the B10 scatter of 0.15~dex throughout our
analysis,  though we test the effect of changing the scatter on our
results in  \S~\ref{scatter}.  We also note that we do not consider
the fact that the stellar mass of a galaxy may actually
\emph{increase} for some time after merging and becoming a satellite,
as it has been recently shown that star formation in active satellites
may continue for several $\Gyr$ \citep[see][]{wetzel_etal11}.  Such
detailed modeling is beyond the scope and intention of this work.

In summary, we take the following steps in Model~1 for assigning final 
stellar masses to an ensemble of halos.
\begin{enumerate}
\item
We use the Z05 model for subhalo evolution to obtain a list of halos 
and subhalos for a range of host halo masses, and we determine the
fractional amount of dark matter mass lost from each subhalo.
\item
We use the B10 stellar-to-halo mass relation to assign stellar masses
to host halos and subhalos at the time of accretion.
\item
For every subhalo, we compute a final stellar mass by relating the 
fractional amount of stellar mass lost to the fractional amount of 
dark matter mass lost through the single free parameter $\epsilon$,
as described in Eqn.~\ref{eqn:main_equation}.
\end{enumerate}


\subsection{Model 2}\label{model_2}


Model~1 operates under the basic assumption that the fractional amount
of satellite galaxy stellar mass loss is proportional to 
subhalo dark matter loss through the free parameter $\epsilon$.  This 
means that even a small amount of dark matter stripping will be 
accompanied by some stellar stripping.  However, as we argued in 
\S~\ref{sec:motivation}, a subhalo may lose a considerable amount of
dark matter before its galaxy is significantly disturbed.  For example, subhalos 
in high-resolution simulations lose a significant fraction of mass at 
their outskirts rapidly upon merging into a larger system, 
while their interiors remain unaltered by this mass loss \citep[e.g.,][]{diemand_etal07}. 
Accordingly, we consider an alternative model that incorporates a 
delay between the initial loss of dark matter mass and stellar 
mass loss.


Our second model states that stellar mass will only be lost if a
sufficient amount of dark matter is lost first.  The model works as
follows.  We first define an approximate \emph{radius of influence}
that a  galaxy has within its subhalo, corresponding to the region
within which the  gravity due to the stellar component is comparable
to that from dark matter.   To estimate this radius, we first assign a
stellar mass to each subhalo at the time of accretion,
$\Mstar_{\mathrm{acc}}$, using the B10 formalism.   Next we assume,
for simplicity, that this is a point mass and we do  not consider the
total amount of ``cold baryons'' or assume a galaxy  profile.  To find the radius
of influence, $\rinfl$, of the stellar mass  associated with the
galaxy, we calculate where the dark matter mass  enclosed within the
subhalo is equal to $\Mstar_{\mathrm{acc}}$.  We thus integrate the
NFW mass profile,
\begin{equation}\label{eqn:enclose_mass}
\MDM_{\mathrm{acc}}(<\rinfl) = 4\pi\rho_{0}r_{s}^{3}\Bigg[\mathrm{ln}\bigg(1 +
\frac{\rinfl}{r_{s}}\bigg) - \frac{\rinfl/r_{s}}{1 +\rinfl/r_{s}}
\Bigg] ,
\end{equation}
where $\rho_0$ and $r_s$ are the NFW parameters, and solve for $\rinfl$
by setting $\MDM_{\mathrm{acc}}(<\rinfl) = \Mstar_{\mathrm{acc}}$.  

The radius $\rinfl$ gives the rough scale within which mass must be lost 
before significant \emph{stellar} mass is lost; however, we allow for 
flexibility in this final prescription for stellar mass loss.  
We assume that a galaxy will start losing stellar mass once
dark matter is stripped from inside a radius that scales linearly 
with $\rinfl$.  We define a new radius, $\rprime = \psi \times \rinfl$, 
where $\psi$ is a free parameter of order unity.  
If the subhalo loses so much dark matter that its
final mass is less than $\MDM_{\mathrm{acc}}(<\rprime)$, we allow
stellar stripping to occur.  Furthermore, we assume that the fraction
of stellar mass lost is equal to the fraction of dark matter lost
\emph{after the bound mass of the subhalo crosses below this threshold}.  
The final stellar mass of the galaxy is thus
\begin{equation}\label{eqn:main_equation_delay_model}
\Mstar_{\mathrm{fin}} =  \Mstar_{\mathrm{acc}}
\left[1 - \mathrm{max}\left\{0,
\left( \frac{\MDM_{\mathrm{acc}}(<\rprime) - \MDM_{\mathrm{fin}}}{\MDM_{\mathrm{acc}}(<\rprime)}\right)\right\}\right],
\end{equation}
which is  analogous to Eqn.~\ref{eqn:main_equation}.

The right-hand side of Figure~\ref{fig:cartoon} illustrates how Model~2 
works.  After accretion, if the subhalo shrinks enough due to mass
loss such that its mass at the final redshift of interest is less 
than $\MDM_{\mathrm{acc}}(<\rprime)$, then stellar mass will be lost
($\Delta \Mstar$) in proportion to the additional amount of dark matter 
lost.  However, if mass loss does not reduce the subhalo 
mass to less than $\MDM_{\mathrm{acc}}(<\rprime)$, no stellar 
mass is lost.  Model~2 incorporates a delay between the 
stripping of the outer layers of dark matter and the stripping of the 
stellar material residing in the depths of the subhalo potential well.  
The single free parameter of Model~2, $\psi$, allows us to vary the 
amount of lag.  As was the case for $\epsilon = 0$ in Model~1, 
setting $\psi=0$ in Model~2 leads to zero stellar mass loss in all subhalos.

In summary, we take the following steps in Model~2 for assigning final
stellar masses to an ensemble of halos.
\begin{enumerate}
\item
We obtain a population of halos and assign stellar masses at the time
of accretion as described in the first two steps of the Model~1 summary.
\item
For each subhalo, we calculate $\rinfl$ using Eqn.~\ref{eqn:enclose_mass}.
We then scale this radius using the free parameter $\psi$, to get
$\rprime = \psi \times \rinfl$.
\item
We calculate the dark matter mass at accretion that is enclosed by 
$\rprime$, $\MDM_{\mathrm{acc}}(<\rprime)$.  If insufficient dark matter
loss has occurred, such that the final subhalo mass is greater than this,
then we assume that no stellar mass loss has taken place.  Otherwise,
we estimate a final stellar mass using Eqn.~\ref{eqn:main_equation_delay_model}.
\end{enumerate}

It is important to note the effect of allowing the stellar and dark
matter components to have distinct density profiles. Using N-body
simulations, \citet{P08} studied the tidal evolution of satellite
galaxies embedded in dark matter halos.  They adopted the
observationally-motivated assumption  that stellar components have
cored (King) profiles, while dark matter follow cuspy NFW models. They
helped explain the extremely large mass-to-light (M/L) ratios of
ultrafaint MW dwarfs by showing that the  cuspy nature of the NFW
profile keeps the central dark matter component more tightly bound
than the stars.   For our Model~2, we have taken $\epsilon = 1$ for
simplicity in order to illustrate the basic features of a simple model
as well as the manner in which it can be constrained. Taking
$\epsilon=1$ in this model allows for  stellar mass loss with a
particular time dependence.  Specifically, M/L ratios can evolve only
when  mass loss does not occur interior to the threshold radius
$\rprime$.  Any additional mass loss interior to $\rprime$ keeps the
M/L ratios fixed.  There are two considerations to this end. First,
allowing $\epsilon$ to take on a value other than 1 would give a
time-dependent M/L ratio even for mass loss interior to $\rprime$.
However, our goal is to  present a simple model and illustrate
constraints on such a model.  A second parameter  would lead to
inevitable degeneracies among parameters and introduce complexities
that would detract from the  broader point of the current paper, and
we show below that a single-parameter model adequately  describes
various observational data.  Second, in even greater detail, we could
consider a  time-varying $\epsilon=\epsilon(t)$, to describe a very
broad range of variations in M/L ratios with time.   Though
not currently necessary to describe the data we consider, such a model
may be more physically appropriate,  and the time variation of
$\epsilon$ could be derived, at least approximately, from specific
assumed stellar mass profiles.   In follow-up work, we are building on
the complexity of the models we present here, though the details of
these broader models are well beyond the scope and intent of this
paper.



\section{CONSTRAINING SATELLITE GALAXY STELLAR MASS LOSS USING GALAXY CLUSTERING}
\label{sec:compare_clustering}


We now turn to constraints on the relationship between stellar mass and 
halo mass imposed by the clustering of galaxies.  Our two stellar mass 
loss models specify the final stellar mass in any halo or subhalo, given a 
value for the parameter $\epsilon$ (for Model~1) or $\psi$ (for Model~2).  
With stellar masses assigned to all halos and subhalos, we can
predict a \emph{halo occupation distribution} 
\citep[HOD: e.g.,][]{peacock00a,scoccimarro01a,berlind02,cooray02}
for any given stellar mass threshold. 
In particular, we compute the mean number of galaxies as a function
of host halo mass, $\langle N \rangle_\mathrm{M}$.  We calculate the
mean over the 500 host halo realizations at each host mass.  
We use this function to compute the number density of galaxies by weighting the 
host halo abundances by $\langle N \rangle_\mathrm{M}$ and integrating over all
halo masses,
\begin{equation}
\bar{n}_{g} = \int_{0}^{\infty}dM\frac{dn}{dM}\langle N \rangle_{M} .
\end{equation}
We adopt the \citet{warren06} halo mass function, $dn/dM$, in this
calculation, though our results are not sensitive to the
specific choice of mass function.  In this way, we find the stellar mass
threshold that yields a galaxy number density equal to that of the 
observed sample with which we aim to compare.

We compare our model predictions to four of the Sloan Digital Sky Survey
\citep[SDSS:][]{york00a} luminosity threshold galaxy samples measured
by \citet{zehavi11}.  Specifically, we consider volume-limited 
samples with $r$-band absolute magnitude thresholds of $M_r \le -18$, 
$-19$, $-20$, and $-21$ (number densities for these samples are listed in Table~2 of
\citealt{zehavi11}).  We do not consider the fact that the observed relationship 
between stellar mass and luminosity is not one-to-one (i.e., galaxies of 
a fixed luminosity can have different stellar masses which, for instance, 
is manifested as scatter in the Tully-Fisher relation -- \citealt{bell01}). 
However, including this scatter should be similar to increasing the scatter 
in the B10 relation between dark matter mass and $\Mstar_{\mathrm{acc}}$, 
which has a very small effect, as we show in \S~\ref{scatter}.

Once we have determined the appropriate stellar mass threshold, we
use the resulting HOD to calculate the predicted clustering of the model.  
We place galaxies within the host halos in a N-body simulation to
measure the 2PCF.  We use a single realization of the ``Consuelo'' 
simulation (with a box size of 
$420 \hmpc$ and $1400^{3}$ particles), which is part of the 
\emph{LasDamas} suite of simulations (McBride et al., in prep.).  
The cosmology assumed in our semi-analytic subhalo model is set to match that of the 
LasDamas simulations\footnote{Throughout the paper, we work within the standard,
vacuum-dominated, cold dark matter ($\Lambda$CDM) cosmological model
with $\Omega_{\mathrm{m}}=0.25$, $\Omega_{\Lambda}=0.75$,
$\Omega_{\mathrm{b}}=0.04$, $h_{0}=0.7$, $\sigma_{8}=0.8$,  and
$n_{\mathrm{s}}=1.0$.} and is similar to the recent WMAP7 values
\citep{komatsu_WMAP711}.  We use the spherical over-density (SO) halo
finder ``Rockstar'' \citep{behroozi_rockstar11} to identify halos.  
We use an SO halo finder because the SO algorithm mimics the assumptions 
made in our semi-analytic model.  Both the 
Z05 model and the halo finder define virial masses based on the 
virial threshold definition of \citet{bryan_norman98}.  We note, however, 
that our results are not sensitive to the choice of halo finder.  We 
have repeated our analysis using a Friends-of-Friends halo finder and 
derived similar results.  We populate host halos 
with galaxies according to our $\langle N\rangle_{M}$ model predictions.  

In order to compare to data, we convert our real-space correlation
functions, $\xir$, to \emph{projected} correlation functions, $\wprp$,
by integrating along lines of sight \citep{davis83,zehavi04a}:
\begin{equation}
\wpp\rp = 2\int_0^{y_{\mathrm{max}}}
\xi\Big(\sqrt{r_{\mathrm{p}}^2+y^2}\Big)dy .
\end{equation}
For each luminosity threshold, we integrate out to the same
$y_{\mathrm{max}}$ as \citet[][listed as $\pimax$ in their
Table~2]{zehavi11}.  We predict $\wprp$ for the four luminosity 
samples, given any value of $\epsilon$ from
Model~1 or $\psi$ from Model~2.  In order to constrain our stellar
mass loss models, we run a grid of $\epsilon$ and $\psi$ values in
steps of $\Delta \epsilon = 0.1$ and  $\Delta \psi = 0.05$.   For each parameter
value and each luminosity threshold,  we compute $\chi^{2}$ by
comparing our model prediction to the \citet{zehavi11}  measurement of
$\wprp$, using only diagonal errors (not the full covariance matrices).  
%


\begin{figure}
\begin{center}
\includegraphics[width=.5\textwidth]{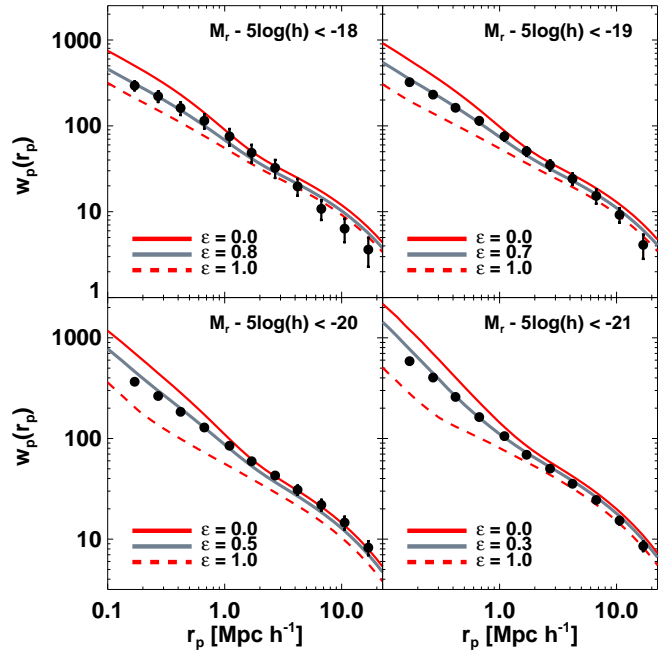}
\caption{Clustering predictions for Model~1.  Model~1 sets the
fractional amount of satellite galaxy stellar mass loss to be
proportional to the fractional amount of subhalo dark matter mass loss
through the free parameter $\epsilon$ (see \S \ref{model_1} for model
details and Fig.~\ref{fig:cartoon} for an illustration).  Each panel
shows results for a different galaxy luminosity threshold.  Solid red
curves show model predictions for $\epsilon = 0$, equivalent to no
stellar mass loss.  Dashed red curves show the $\epsilon = 1$ case,
where the stellar mass loss occurs at the same rate as subhalo dark
matter mass loss.  Grey curves show the $\epsilon$ values that match 
the observed SDSS clustering measurements of \citet[][black points]{zehavi11}.}
\label{fig:model1_comparison}
\end{center}
\end{figure}



\begin{figure}
\begin{center}
\includegraphics[width=.5\textwidth]{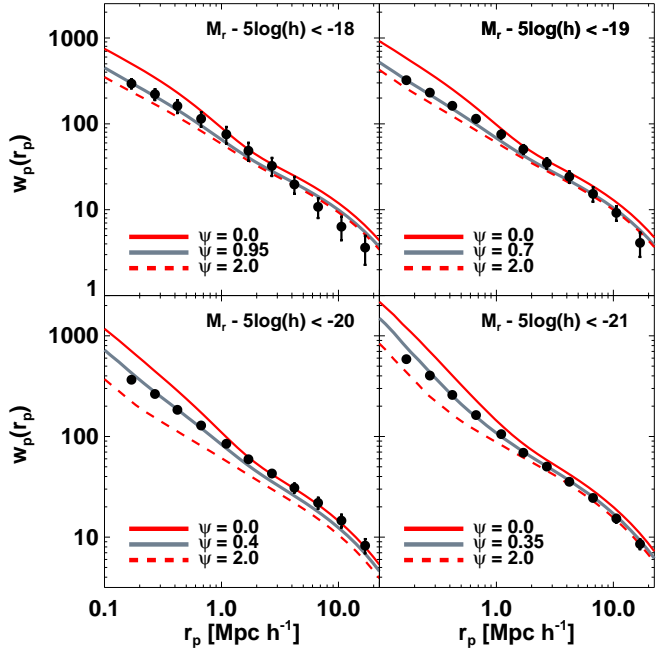}
\caption{Clustering results for Model~2.  Model~2 defines a radius of
influence, $\rinfl$, for a galaxy within a subhalo.  If a subhalo
loses sufficient dark matter mass such that it starts to lose mass
inside of $\rinfl$, then stellar mass is lost at a rate proportional
to the subsequent subhalo mass loss.   The free parameter $\psi$
allows $\rinfl$ to grow or shrink, thus a larger value of $\psi$ means
that more stellar mass loss will occur on average for  satellite
galaxies.  Solid red curves show model predictions for $\psi = 0$,
equivalent to no stellar mass loss, thus the same as $\epsilon = 0$
for Model~1.  Dashed red curves show the predictions for $\psi = 2$,
an arbitrary value chosen to represent how too much stellar mass
loss will under-predict $\wprp$. Grey curves show the $\psi$ values
that match the observed  clustering for the same luminosity threshold
samples as in Fig.~\ref{fig:model1_comparison}.}
\label{fig:model2_comparison}
\end{center}
\end{figure}


Figure~\ref{fig:model1_comparison} shows clustering results for
Model~1 compared to measurements from \citet{zehavi11} for our four
SDSS luminosity threshold samples: $M_{r} < -18, -19, -20 \text{ and }
-21$.  To reiterate, Model~1 assumes that the fractional amount of
satellite galaxy stellar mass loss is proportional to the fractional
amount of subhalo dark matter mass loss with proportionality constant 
$\epsilon$.  Adopting $\epsilon = 0$ specifies no stellar mass loss and 
is equivalent to using $\Vmax$ at the epoch of accretion for abundance matching.  
On the other hand, adopting $\epsilon =1$ means that stellar mass is lost 
at the same rate as dark matter.  For each
luminosity sample, we show the $\epsilon = 0$  and~1 cases as red
solid and dashed lines, respectively.  

At each luminosity,  $\epsilon = 0$ predicts small-scale
clustering that is stronger than the data indicate and this 
gets to the essence of using clustering to constrain stellar mass loss.  
This over-prediction can be attributed to the fact that no stellar mass loss 
has occurred for satellite galaxies, resulting in too many satellites with 
high stellar masses.  This leads to enhanced clustering on small scales
\citep[e.g.][]{watson_powerlaw11}.  To be consistent with the data,
some stellar  mass loss needs to occur.  However, the $\epsilon = 1$
result demonstrates that  too much stellar mass loss leads to weaker clustering 
than the data require.  The grey curves  in
Figure~\ref{fig:model1_comparison} correspond to intermediate 
amounts of stellar mass loss and represent the $\epsilon$ 
with the lowest $\chi^{2}$ values.\footnote{We note that none of our
model  predictions match the large-scale clustering for the $M_{r} <
-18$ sample.   This may be attributed to the finite volume  of the
sample, as discussed in \S~3.2 of \citet{zehavi11}.}  Despite the fact
that Model~1 paints a simplified picture of how stars  are stripped
from the galaxies they reside in, we show that it is very  effective
at matching the observed clustering.  Moreover, it is striking  that
$\epsilon$ decreases with increasing luminosity.  We will discuss
this luminosity dependence in detail in \S~\ref{sec:lum_dependence}.

Figure~\ref{fig:model2_comparison} shows similar results for Model~2.
This model was designed to allow dark matter on the periphery of a 
subhalo to be lost due to the strong tidal field of the host, without 
significantly altering the luminous galaxy residing deep in the core of the subhalo.  
Model~2 mimics this ``lag'' by defining a critical radius within each subhalo and 
only allowing for stellar mass loss if the subhalo loses mass from within that 
radius.  This radius can be varied through the parameter  $\psi$.  The 
limit $\psi = 0$ corresponds to no stellar mass loss and is equivalent 
to abundance matching using $\Vmax$ at accretion.  
In Model~1, $\epsilon =1$ is analogous to using the 
subhalo mass at the final redshift output for abundance matching.
In Model~2, this happens when the critical radius is larger than the 
size of the subhalo at accretion.  This occurs at a different value of 
$\psi$ for each subhalo.  Thus, we choose an arbitrary value $\psi = 2$ 
to again illustrate how too much stellar mass loss will under-predict 
clustering.  
%
As was the case for Model~1, we are able to match the observed 
clustering and we see a strong evolution in $\psi$ as a function of 
luminosity.  This emphasizes what was found for Model~1, that low-luminosity
galaxies experience more efficient stellar stripping throughout their 
evolution in a host halo than luminous galaxies.


\subsection{The Effect of Scatter in the Stellar-to-Halo Mass Relation}\label{scatter}


%

Observations indicate that halos of a given mass can host galaxies
with a range of luminosities and stellar masses
\citep[e.g.,][]{vdBosch07,zheng07,yang08,more09b}.  For this reason, it 
is overly restrictive to assume a one-to-one relation between halo mass
and stellar mass, whereas a relationship with some intrinsic
scatter is more appropriate.  This is especially true for
abundance matching because the presence of scatter changes the rank
order of galaxies.  We rank our halos and subhalos by stellar mass, so 
the rank order will be affected by the assumed amount of scatter.  Scatter is
typically accounted for by assuming  a distribution of stellar mass at
fixed halo mass, with the mean value of  stellar mass given by a
particular SHMR.  For example, the B10 scatter that  we adopt in this
paper is a log-normal distribution with a dispersion $\sigma=0.15$.
We implement this scatter by randomly drawing from this distribution
when we  assign stellar masses to halos and subhalos at the time of
accretion.  The  results shown in Figures~\ref{fig:model1_comparison}
and~\ref{fig:model2_comparison}  include this scatter.


\begin{figure}
\begin{center}
\includegraphics[width=.5\textwidth]{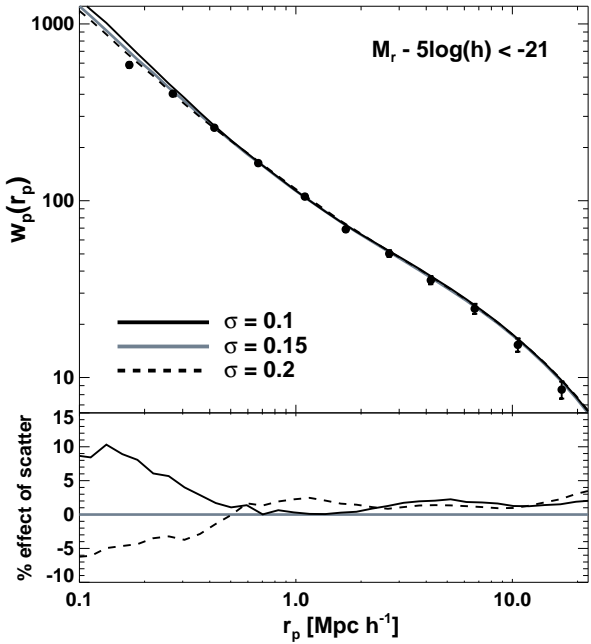}
\caption{
The effect of the scatter in the stellar mass -- halo mass relation,
on the clustering of a stellar mass threshold galaxy sample. \emph{Top panel}: 
The grey curve shows the predicted correlation function from Model~1 with 
$\epsilon = 0.3$ and our assumed scatter of $\sigma = 0.15$ dex, given by
\citet{behroozi10}.  Solid and dashed black curves show results for changing
the scatter to $\sigma = 0.1$ dex and 0.2 dex, respectively.  Each curve is
averaged over three realizations of that amount of scatter.
\emph{Bottom panel}: The percent deviations in the correlation function from
the fiducial scatter of $\sigma=0.15$.  At the scale of the innermost data 
point of \citet{zehavi11} that we compare our model results to ($\sim 0.17 \hmpc$), 
lowering the scatter to $\sigma =0.1$ increases $\wprp$ by $\sim 8\%$
and increasing the scatter to $\sigma=0.2$ lowers $\wprp$ by $\sim 5\%$.
On larger scales the effect is much smaller.
}
\label{fig:scatter_effect}
\end{center}
\end{figure}



\begin{figure*}
\begin{center}
\includegraphics[width=.85\textwidth]{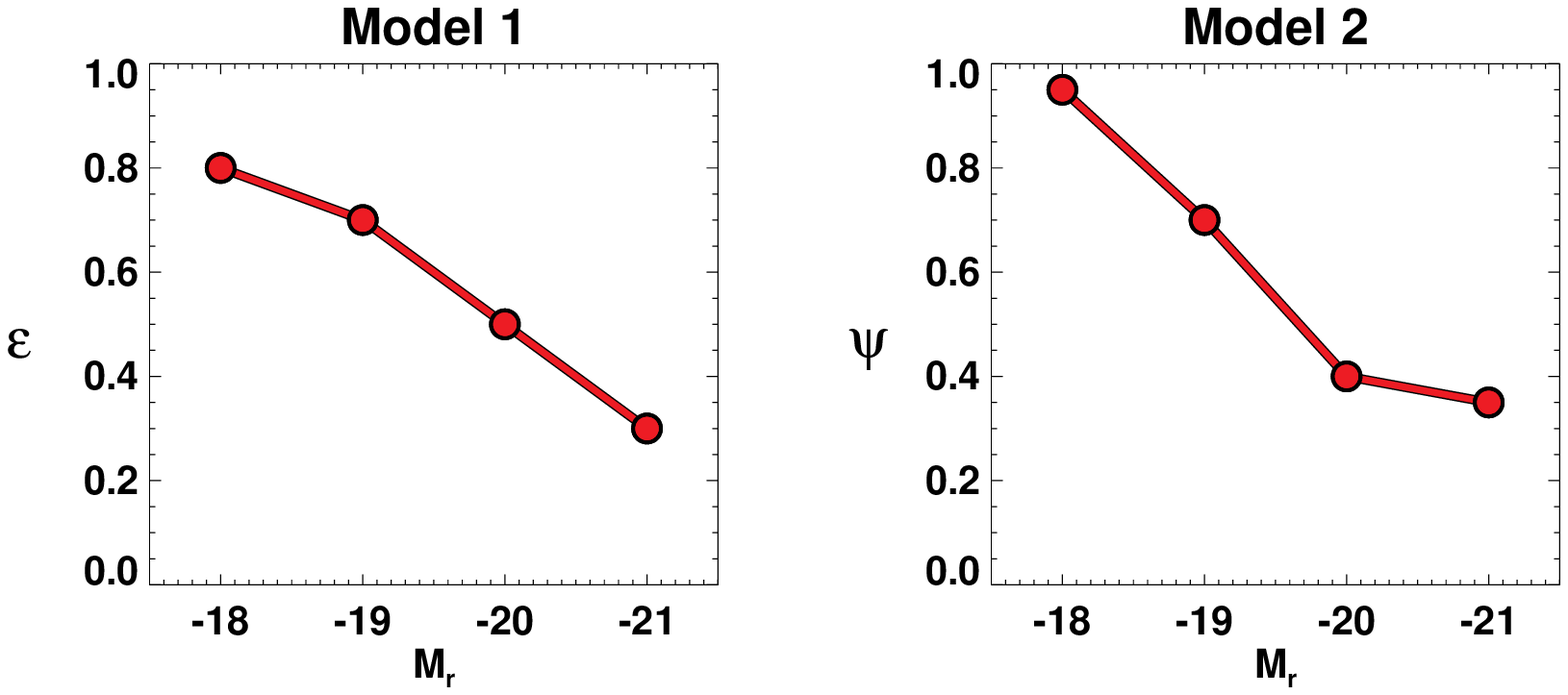}
\caption{The luminosity dependence of satellite galaxy stellar mass
loss.  \emph{Left panel}: The best-fit Model~1 parameter $\epsilon$ as a
function of absolute $r$-band magnitude.  The decreasing trend of $\epsilon$ 
with increasing luminosity means that low-luminosity satellite galaxies experience
greater stellar mass loss relative to subhalo dark matter mass loss than 
luminous galaxies.  \emph{Right panel}: The best-fit Model~2 parameter $\psi$ 
as a function of absolute $r$-band magnitude.  The decreasing trend of $\psi$
with increasing luminosity means that low-luminosity satellite galaxies have a 
greater radius of influence (causing more stellar mass loss) than luminous galaxies.  
In other words, the same qualitative luminosity trend is seen in both models.}
\label{fig:epsilon_psi_vs_luminosity}
\end{center}
\end{figure*}


We repeat our analysis with different dispersions in order to test
the sensitivity of our results to uncertainty of the scatter in the SHMR.
Figure~\ref{fig:scatter_effect} Shows the effect of changing the SHMR 
dispersion for one of the models that matches the clustering of 
$M_{r} < -21$ galaxies:~Model~1 with $\epsilon=0.3$.  This sample is the 
most sensitive to scatter because the slope of the SHMR is shallowest
for luminous galaxies and thus a given amount of scatter in  stellar
mass at fixed halo mass translates into a large scatter in halo mass
at  a fixed stellar mass threshold (see Fig.~\ref{fig:SHMR_Behroozi}).  
The top panel of Figure~\ref{fig:scatter_effect} shows the correlation 
function for three SHMR dispersions: $\sigma=0.1$, 0.15, and 0.2.   
The bottom panel depicts percent deviations from the
fiducial  scatter of $\sigma=0.15$.  This range is conservative, as it
is larger than the  errors in scatter quoted in B10 (especially at the
low end).  The figure  demonstrates that changing the scatter has a
small effect on $\wprp$, in the  sense that more scatter generally
leads to a lower amplitude of clustering on  small scales.  Over the
range of data points from the \citet{zehavi11} data  that we model,
there is a maximum $\sim 8\%$ and $\sim 5\%$ shift in $\wprp$  for the
$\sigma=0.1$ and $\sigma=0.2$ cases, respectively.  This is not a
large effect and does not result in a significant change to the
best-fit values of  $\epsilon$ or $\psi$.  
%
Uncertainties in the scatter of the  SHMR do not significantly alter
our results or our primary conclusions.


\section{Luminosity Dependence of Satellite Galaxy Stellar Mass Loss}\label{sec:lum_dependence}



\begin{figure*}
\begin{center}
\includegraphics[width=1.\textwidth]{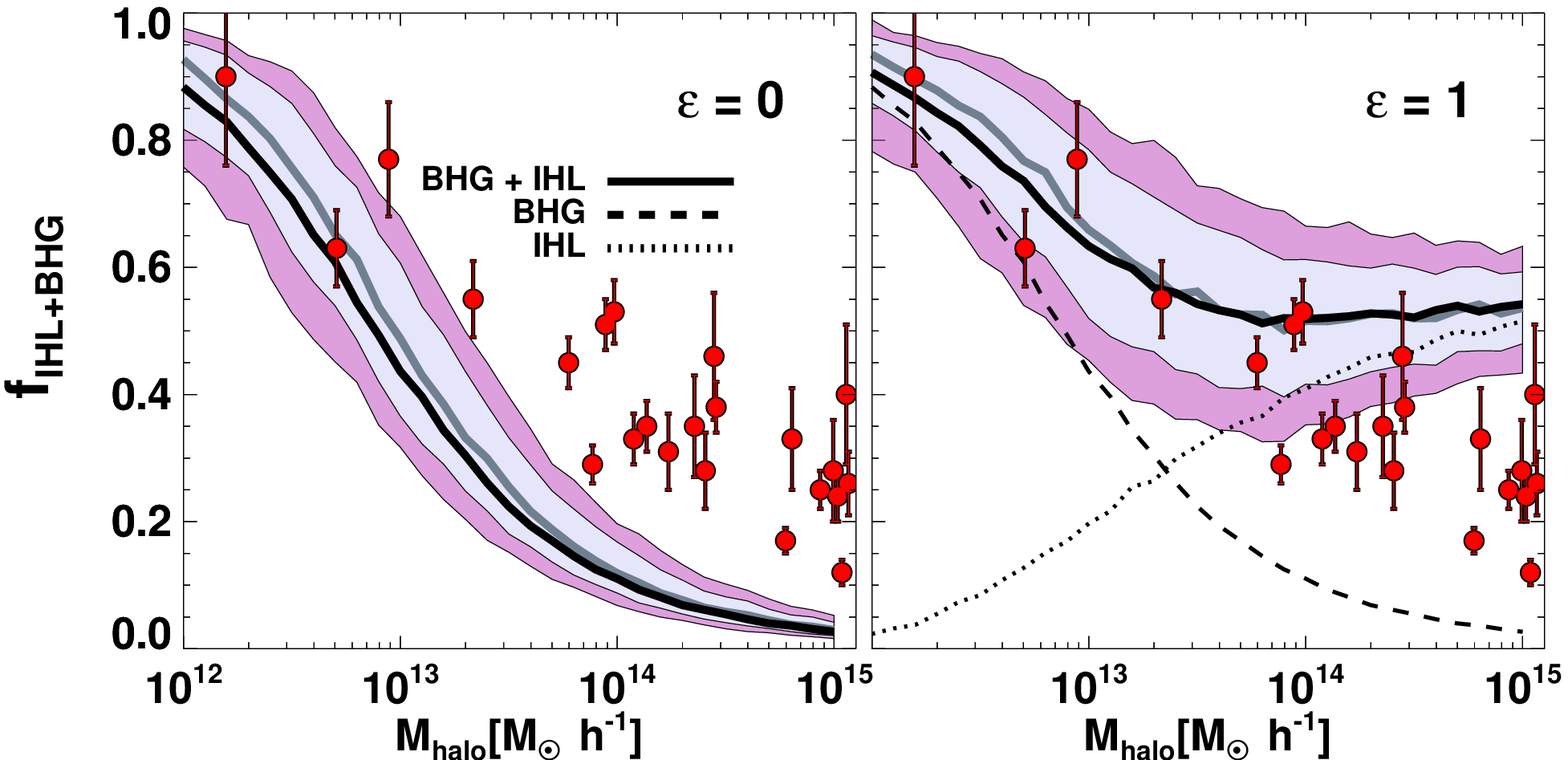}
\caption{$\fihlbhg$ predictions for the two extreme cases of Model~1:
$\epsilon = 0$ and 1.  $\fihlbhg$ is the combined amount of stellar
mass  from the brightest halo galaxy (BHG) and the intrahalo light
(IHL) divided by the total stellar mass of a system (the BHG, IHL, and
the stellar mass still associated with surviving subhalos).  Thick solid
black curves show the mean $\fihlbhg$ as a function of host halo mass and 
are computed from 500 model realizations of each host mass, while
grey curves represent the median.  Light and dark  purple shaded
regions represent the $68\%$ and $95\%$ range of the model
distributions.  BHG and IHL contributions are shown individually as
dashed  and dotted curves, respectively.  Red points show the
\citet{gonzalez07} data measurements for comparison.  \emph{Left
panel}: $\epsilon = 0$ means that no stellar mass loss occurs,
resulting in no IHL.  As a result, $\fihlbhg$ is strongly
under-predicted  at cluster scales ($M_\mathrm{{halo}}\sim 14-15
\hMsun$) relative to the data.   \emph{Right panel}: $\epsilon = 1$
means that stellar mass loss occurs at  the same rate as subhalo dark
matter mass loss.  This results in an over-prediction of the IHL.
Note: data points are shifted  by $\sim 40\%$ in order to convert
the $M_{500}$ ($\Delta_{\mathrm{crit}}=500$)  masses given in
\citet{gonzalez07} to our virial masses (using the definition $\Delta_{\mathrm{vir}} = 377$).}
\label{fig:fIHL_eps0_eps1}
\end{center}
\end{figure*}



\begin{figure*}
\begin{center}
\includegraphics[width=1.\textwidth]{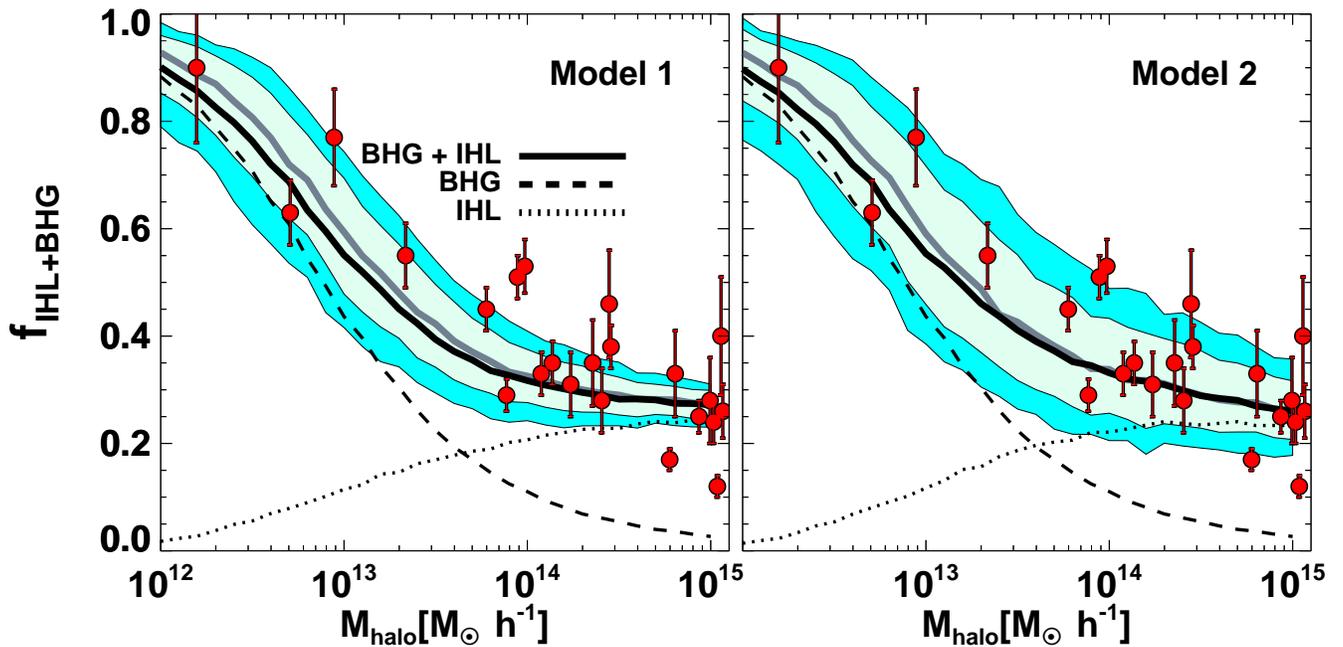}
\caption{$\fihlbhg$ predictions for our best-fit Model~1 (left panel)
and  Model~2 (right panel).   We compute each subhalo's contribution
to the  IHL according to its initial stellar mass by using the
appropriate best-fit  values of $\epsilon$ or $\psi$.  For example, a
subhalo with a stellar mass  that corresponds to a galaxy of absolute
$r$-band magnitude equal to -18.5  loses stellar mass according to
$\epsilon=0.8$ (in Model~1), whereas a  subhalo with a stellar mass
corresponding to $M_r=-20.5$ loses stellar mass according to
$\epsilon=0.5$.  Shaded regions, curves, and points are as in
Fig.~\ref{fig:fIHL_eps0_eps1}.  Our stellar mass loss models predict
that the mean IHL fraction in halos  increases from a few
percent at the scale of individual galaxies  ($M_\mathrm{{halo}}\sim
10^{12}\hMsun$) to $\sim 20 - 25\%$ at cluster scales
($M_\mathrm{{halo}}\sim 10^{14}-10^{15}\hMsun$).  Both models succeed in
matching the  mean trend observed by \citet{gonzalez07}.  However,
Model~1 seems to  under-predict the scatter in $\fihlbhg$, whereas
Model~2 predicts a scatter  that is more consistent with the data
measurements.  We thus find that this more physically-motivated model
for satellite galaxy stellar mass loss is favored over Model~1.}
\label{fig:fIHL}
\end{center}
\end{figure*}



\begin{figure}
\begin{center}
\includegraphics[width=.5\textwidth]{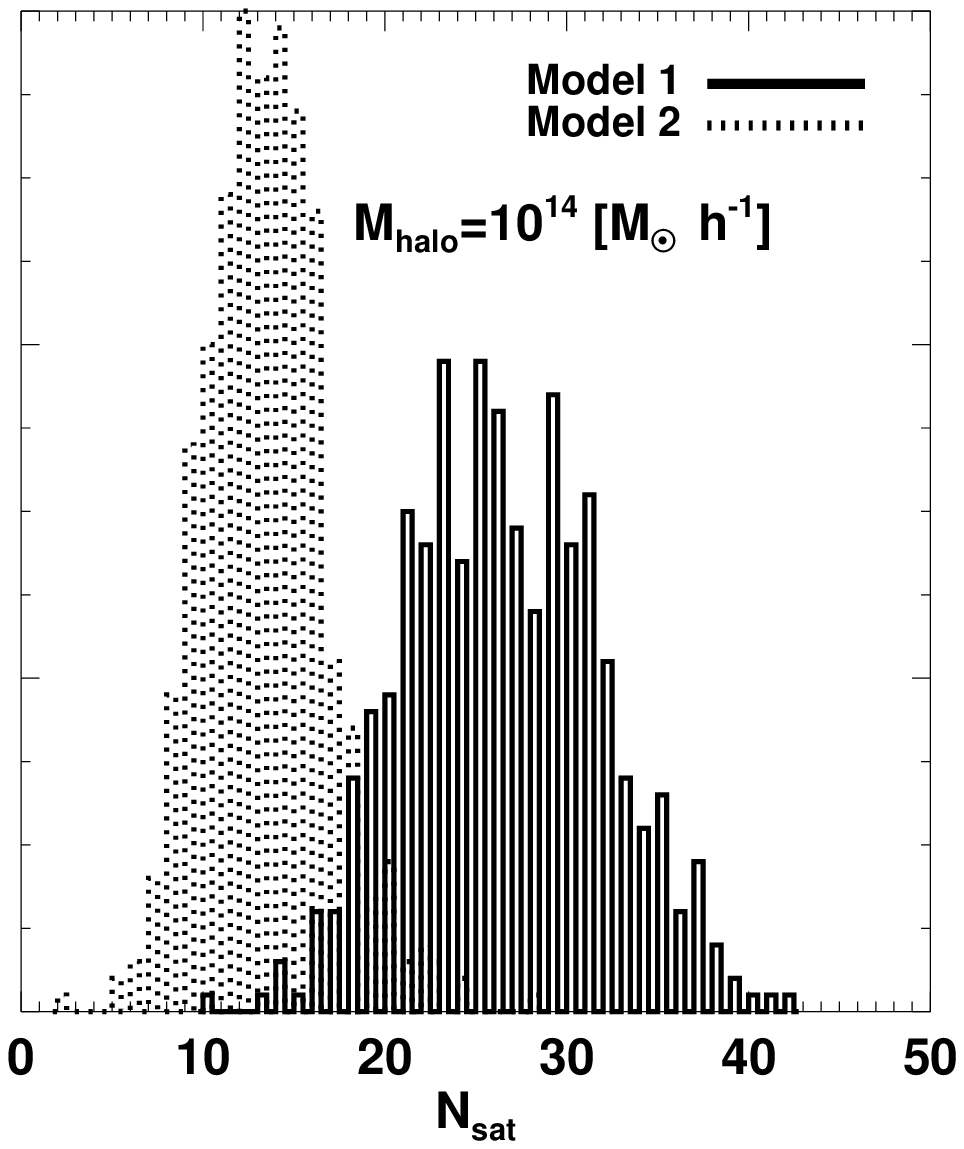}
\caption{Histograms of the number of satellites that contribute $75\%$
of the total IHL to a host halo of mass $10^{14}\hMsun$ for 500
realizations of Model~1 (solid histogram) and Model~2 (dotted
histogram).  It is clear that in Model~2
the same amount of IHL comes from a much smaller number of satellites
that have each been stripped of more stars than in Model~1.  Since fewer
satellites contribute to the IHL in Model~2, the Poisson scatter in
this model is larger and is evident in the $\fihlbhg$ scatter in
Fig.~\ref{fig:fIHL}}.
\label{fig:scatter}
\end{center}
\end{figure}


Figures~\ref{fig:model1_comparison} \&~\ref{fig:model2_comparison} 
indicate that the best-fit values of both $\epsilon$ and $\psi$ 
vary with luminosity.  We highlight these trends in 
Figure~\ref{fig:epsilon_psi_vs_luminosity}.  The left-hand panel 
summarizes results from Model~1, and demonstrates that the best-fit value 
of $\epsilon$ drops from $\epsilon=0.8$ for $M_{r} < -18$ galaxies 
to $\epsilon=0.3$ for $M_{r} < -21$ galaxies.  There is a strong luminosity dependence 
associated with satellite galaxy stellar mass loss, such that relatively 
low-luminosity satellite galaxies lose more stellar material per unit of 
dark matter than more luminous satellite galaxies.  The right-hand panel of 
Figure~\ref{fig:epsilon_psi_vs_luminosity} summarizes results from Model~2, 
which show a consistent trend.  The best-fit value of $\psi$ drops from $\psi=0.95$ for 
$M_{r} < -18$ galaxies to $\psi=0.35$ for 
$M_{r} < -21$ galaxies.  Lower values of $\psi$ indicate
less stellar mass loss for a given amount of dark matter mass loss.  
The two models we explore absorb a significant number of subtle 
effects into simple assumptions.  Model~1 and Model~2 differ substantially in detail, 
yet both analyses indicate that the luminosity dependence of galaxy 
clustering can be explained if low-luminosity satellite galaxies 
lose stars more efficiently than luminous galaxies.

What does this result mean?  Presumably it means that the stars within
luminous galaxies are more tightly bound -- relative to their surrounding
dark matter -- than the stars in less luminous galaxies.  This could be
due to luminous galaxies having more compact stellar density profiles.  
In fact, it may be that this result can be explained by the well-known
correlation between galaxy luminosity and morphology, whereby luminous 
galaxies are more likely to be ellipticals than low-luminosity galaxies.
For example, \citet{blanton03} showed that the mean S\'ercic index 
of galaxies roughly doubles from $M_r=-18$ to $M_r=-21$ (looking at their
Fig.~9 and converting $i$-band to $r$-band magnitudes).  We must, of 
course, keep in mind that these morphology-luminosity correlations
apply to all galaxies, of which satellites are only a small portion.
It may be that these correlations would vanish in samples containing
only satellite galaxies.  

An ancillary consideration is the relative size of galaxies compared
to  the typical (sub)halos they occupy.  For instance, using a sample
of 140,000 SDSS galaxies, \citet{shen_etal03} showed the size distribution
(half-light radii) of early- and late-type galaxies as a function of
$r$-band luminosity and stellar mass.  Examining their Figs.~2 \&~3, the
half-light radius $\overline{R}$ increases roughly by a factor of $\sim3$ from $M_{r} <
-19.25$ to $-22.25$ for late-type galaxies, and by a factor of $\sim5$
for early types.  However, if we consider the typical halo masses that
host these galaxies (see $\Mmin$ values for $M_{r} < -19$ and $-22$
from Table~3 of \citealt{zehavi11}), they correspond to virial radii
of $\sim 125\hkpc$ and $\sim 750\hkpc$, respectively, a factor of 6.
In other words, at a very approximate level, larger (more luminous)
galaxies are slightly deeper within their host halo potential wells
than smaller (less luminous) galaxies, and are relatively 
less susceptible to stellar stripping.

The luminosity trend that we have found demonstrates the power of our 
modeling approach.  We use one observational measurement, galaxy 
clustering, to shed light on an important physical process that is 
essentially unobservable: stellar mass loss from satellite galaxies 
relative to dark matter mass loss.  In the next section we show how our 
models can be used to predict a different observable, the IHL.  The 
IHL is a remnant of stellar mass loss, so it may be used to probe the 
liberated stellar debris, and thus provide an important cross-check on 
our results.


\section{PREDICTIONS FOR INTRAHALO LIGHT AT VARYING SCALES}\label{sec:predict_IHL}


If stars are freed from their host galaxies they will presumably
become part of the IHL.  Some of the IHL is a diffuse background,
while some may be composed of coherent streams, a prominent example of 
which  is the Sagittarius Stream in our own galaxy 
\citep[e.g.,][]{ibata01,dohm_palmer01,newberg02}, if the stars were 
stripped from a subhalo in the past few dynamical times.
Observations of the IHL are notoriously difficult, and
\citet{gonzalez07}  argue that the proper quantity to measure is the
sum of the light from the host central galaxy (the ``brightest halo
galaxy'' or BHG) and the diffuse intrahalo light component, because it
is non-trivial to disentangle the two.  Our models assign stellar
masses to all halos (both hosts and subhalos), so we can compute the
contribution of the BHG to the total stellar mass of the system.  This
combination is also less subject to theoretical uncertainty  because
collisions between the BHG and an infalling satellite can  disperse
stellar mass into the IHL.  While parsing of the stellar mass  of the
BHG and IHL individually is sensitive to the fraction of  stellar mass
ejected in such collisions, the sum is robust to  uncertainties in
this process.

We start by investigating the predictions for the $\epsilon=0$ and
$\epsilon=1$  cases from Model~1. These extremes were shown in
\S~\ref{sec:compare_clustering} to over- and  under-predict the
observed clustering, respectively, for each luminosity threshold.  In
Figure~\ref{fig:fIHL_eps0_eps1}, we show the fraction of the total
stellar mass of a system (including stellar mass in satellite
galaxies),  that is contained within both the IHL and the BHG,
$\fihlbhg$, as a function  of host halo mass, $M_{\rm halo}$.  The
thick solid black curves in Figure~\ref{fig:fIHL_eps0_eps1} show the 
mean value of $\fihlbhg$ obtained from our 500 realizations of host 
halos, and grey curves represent the median.  The inner and outer 
shaded regions represent the  $68\%$ and $95\%$ ranges of the model 
distributions.  The mean individual BHG and IHL contributions are 
shown distinctly as dashed and dotted curves, respectively.  The
$\fihlbhg$ data from \citet{gonzalez07} are shown as red circles 
with errorbars, and are the same in both panels.

For $\epsilon=0$ there is no stellar mass loss, and thus no
contribution to the total stellar mass of the system from the IHL.
The stellar mass of the BHG alone provides a decent description of the
data for halo masses below group mass scales, $M_{\rm halo} \lesssim
10^{13}\, \hMsun$.  While less massive systems have been observed to
have some detectable IHL (see  \S~\ref{intro}), it is mainly large group- 
and cluster-size objects that are known to have significant IHL components.  
It is therefore not surprising that the $\epsilon=0$ model underestimates 
$\fihlbhg$ on these scales. 

At the other extreme, the $\epsilon=1$ model assumes that stellar
mass is lost with the same efficiency as dark matter.  We argued in
\S~\ref{sec:stripping_models} that this over-estimates  stellar mass
loss, so it is not surprising that choosing  $\epsilon=1$ leads to an
overestimate of $\fihlbhg$ for systems greater than large groups 
($M_{\rm halo} \gtrsim 10^{13.5}\, \hMsun$).  At lower masses, typical 
infalling satellites have halo masses below  
$\MDM_{\rm acc} \lesssim 10^{11}\, \hMsun$ and
bring with them  only a small amount of stellar mass (B10). This is
the regime of the SHMR where stellar mass rapidly decreases with
decreasing halo mass (see Fig.~\ref{fig:SHMR_Behroozi}).  Data are not
yet precise  enough to be sensitive to the stellar mass carried into
the  systems by such small, infalling subhalos.  Consequently,
$\fihlbhg \approx f_{\rm BHG}$ and can be adequately described in
low-mass systems at both the $\epsilon=1$ and $\epsilon=0$ extremes.

We note that in our model predictions for the IHL, we only consider
the  stellar mass stripped from satellite galaxies that have stellar
masses  equivalent to $M_{r} < -18$ galaxies and brighter.  We ignore
stellar mass  loss from less luminous (less massive) satellites in the
$\fihlbhg$  calculation.  However, the stellar mass locked up in these
lower mass  ``uncounted'' subhalos only amounts to $\sim 3\%$ of the
total stellar mass  within a given host halo, so neglecting them has a
negligible effect on our  $\fihlbhg$ model predictions.  We also note
that in Figure~\ref{fig:fIHL_eps0_eps1} we have shifted the data
points to higher masses by $\sim 40\%$ in order  to convert the
$M_{500}$ ($\Delta_{\mathrm{crit}}=500$) masses given in
\citet{gonzalez07} to our virial masses (using the definition
$\Delta_{\mathrm{vir}}=377$).   We assumed an NFW profile to
perform this conversion.

Having shown that these two extreme cases are not able to describe the
data accurately, we turn to the predictions of our models.  To make $\fihlbhg$
predictions at a given host halo mass scale we need to assign each
(sub)halo an $\epsilon$ value for Model~1 and a $\psi$ value for
Model~2.  Let us examine just the Model~1 case for simplicity.  When
we found the best-fit value of $\epsilon$ for each luminosity
threshold sample in \S~\ref{sec:compare_clustering}, we also
determined a final stellar mass threshold for the sample through
abundance matching.  For example, the $M_r<-20$ sample was found to
correspond to halos and subhalos with final stellar masses greater
than some value $\Mstar_{\mathrm{fin},20}$, where final stellar masses
are predicted using $\epsilon=0.5$.  The $M_r<-21$ sample corresponds to
stellar masses greater than $\Mstar_{\mathrm{fin},21}$, where final
stellar  masses are predicted using $\epsilon=0.3$, and so on.  We assign
each subhalo a value of $\epsilon$ using these stellar mass
thresholds.  For example, if a subhalo has a stellar mass at accretion
that falls between $\Mstar_{\mathrm{fin},20}$ and
$\Mstar_{\mathrm{fin},21}$, we assign it $\epsilon=0.5$.  If, instead,
it has a mass between  $\Mstar_{\mathrm{fin},19}$ and
$\Mstar_{\mathrm{fin},20}$, we assign it $\epsilon=0.7$. It is awkward
to use the stellar mass at accretion to assign subhalos to ranges in
$\Mstar_{\mathrm{fin}}$, but we cannot estimate final stellar masses
without having a value of $\epsilon$ in the first  place, so we are
stuck with this approximate method.  We follow the same process for
Model~2 by choosing the appropriate $\psi$ values.  We emphasize that
these $\epsilon$ and $\psi$ values are not ``tuned'' to agree with the
$\fihlbhg$ observations, rather are genuine predictions that result
from parameter values found from matching to clustering data.

Figure \ref{fig:fIHL} shows the Model~1 (left panel) and Model~2
(right panel) $\fihlbhg$ predictions, as well as the individual IHL
and BHG contributions.  While $\fihlbhg$ rapidly decreases with
increasing host halo mass, the contribution to $\fihlbhg$ from the
IHL increases. This supports a picture in which galaxy-mass 
halos have very few luminous satellite galaxies and little IHL, 
so the stellar masses of these systems are dominated by the BHGs.  
Galaxy size halos accrete the majority of their mass from small 
subhalos that have very high mass-to-light ratios \citep{purcell_etal07}.  
Therefore, shredded satellites deposit very little stellar mass into
the IHL.  The number of large satellites increases with host halo mass.  
The larger and more common subhalos within larger host halos may form 
stars more efficiently (have lower mass-to-light ratios) and provide a 
source for developing significant IHL.  Both models find that host masses 
of $M_{\rm host}/\hMsun = 10^{12}, 10^{13}, 10^{14}$, and $10^{15}$, 
contain mean IHL fractions of $\fihl \sim 2\%, 10\%, 20\%$, and $25\%$, 
respectively.  This is in good agreement with several previous studies, 
as we discuss in \S~\ref{sec:conclusion}.  The IHL thus provides
an independent check of our stellar mass loss results.

Figure \ref{fig:fIHL} contains a comparison between our model
predictions and the \citet{gonzalez07} data.  Both models predict mean
$\fihlbhg$ fractions that are consistent with the observations.\footnote{
We note that this discussion assumes that the \citet{gonzalez07} data are
accurate, whereas they may be systematically overestimated due to an
assumed constant mass-to-light ratio \citep{leauthaud11c}.   
}
Interestingly, we find that the scatter predicted by Model~1 is
substantially smaller than it is for Model~2 and it appears to be
inconsistent with the scatter found by \citet{gonzalez07}.
On the other hand, Model~2 is more successful in reproducing the observed 
scatter in $\fihlbhg$.  To quantify this, we calculate the probability 
that the scatter in the \citet{gonzalez07} data points could result from 
the model distributions.  We do this as follows.  First, we restrict the 
comparison to the high mass regime (the last 8 data points) because we 
wish to investigate the scatter at fixed mass (if we include lower mass
points, the measured scatter will be affected by the mean trend of
$\fihlbhg$ vs. mass).  We next create a model data set by choosing 
a single model realization of a host halo (out of the 500) for each 
\citet{gonzalez07} data point, at the appropriate halo masses.  This leads
to 8 model values of $\fihlbhg$.  We then add a Gaussian random error
to these values, using the corresponding \citet{gonzalez07} errorbars.
To fully sample the distribution of data sets that could arise from
our models, we repeat this process 10,000 times.  Finally, we measure the
weighted standard deviation of $\fihlbhg$ values for the observed data, as 
well as each of the 10,000 model data sets.  We find that in Model~1,
only 0.8\% of the model data sets have a scatter that exceeds that of the 
\citet{gonzalez07} data, whereas in Model~2, 25\% of the model data
sets have a larger scatter.  

This statistical test indicates that Model~1 should be rejected (at
the  2.7$\sigma$ level) in favor of Model~2, an indication that is not
entirely surprising given that Model~2 accounts for delayed stellar
mass loss that is to be expected on physical grounds.  Model~1 assumes
that stellar mass loss is simply proportional to subhalo mass loss, so
the scatter predicted by this model must be primarily caused by the
scatter in the total amount of dark matter mass loss, with a secondary
amount of scatter coming from the SHMR.  However, the
"all or nothing" action provided by the radius of influence in
Model~2 somehow amplifies this scatter.  Upon closer
investigation, we find that the difference in scatter between the two
models arises from simple Poisson statistics.  Figure \ref{fig:scatter} shows the
distribution of the number of satellites that contribute $75\%$ of the
IHL to a host halo of mass $10^{14} \hMsun$, measured from 500
realizations for both Model~1 (solid histogram) and Model~2 (dotted histogram).  It is clear that in Model~2
the same amount of IHL comes from a much smaller number of satellites
that have each been stripped of more stars than in Model~1.  This is
a direct result of the nature of Model~2: most subhalos do not lose
stars, but if they do, it is at a high rate.  Since fewer
satellites contribute to the IHL in Model~2, the Poisson scatter in
this model is larger and is manifested in the greater $\fihlbhg$ scatter
 seen in Fig. \ref{fig:fIHL}.


\section{SUMMARY \& DISCUSSION}\label{sec:conclusion}


While galaxy formation is complex, we have shown in this paper that an
understanding of the fates of satellite galaxies provides key insight
into the galaxy formation puzzle. Satellite galaxies live extremely
tumultuous lives.  Their spatial clustering can shed light on how
they lose stellar mass and contribute to building the IHL over a large
range of host halo mass scales. Part of our motivation for using 
galaxy clustering to constrain stellar mass loss stems from possible 
shortcomings in the commonly-employed abundance matching 
technique \citep[e.g.,][Reddick et al., in prep.]{wetzel_white10,yang11b},
which suggest natural generalizations with the gross form of the particular 
models we study.  The standard method of abundance matching
involves mapping galaxies to halos by assuming a 
relationship between halo (and subhalo) mass (or circular velocity) 
and a galaxy property (luminosity or stellar mass) at a particular 
time.  A galaxy luminosity assignment according to halo mass \emph{at 
the time the halo was first accreted onto a larger system} provides 
a very useful description of known galaxy clustering properties \citep[e.g.,][]{conroy06}.  
However, this type of assignment implies that \emph{no
stellar mass loss will occur}.  This is contrary to many 
aspects of galaxy, group, and cluster evolution, including 
the prevailing paradigm in which the IHL is produced by stellar 
mass liberated from infalling galaxies by interactions in the 
group environment \citep[e.g.,][]{conroy07,purcell_etal07,yang09b}.  
Abundance matching is also sensitive to
the resolution of the cosmological simulation on which it is 
implemented.  Subhalos are artificially destroyed by falling below 
the numerical resolution limit and immediately removed from the 
galaxy population irrespective of their circular velocities 
at their accretion times.  This can lead to differing predictions 
for the small-scale clustering of galaxies of a particular luminosity 
among simulations with a fixed mapping of halo size onto luminosity.  

These considerations persuaded us to explore simple models that
associate stellar masses to subhalo masses  and constrain satellite
galaxy \emph{stellar mass loss} by matching to clustering data.  We
implemented two distinct models. Model 1 related the fractional amount
of stellar mass loss to that of the subhalo dark matter mass loss from the
time of accretion until the final redshift through a single free
parameter $\epsilon$.  Model 2 was introduced to mimic a ``lag'' in
stellar mass loss wherein the satellite galaxy at the core of the
subhalo experiences no stellar mass loss until the subhalo has been
stripped of sufficient dark matter on its periphery.   Stellar mass
loss in this model only occurs once dark matter  has been lost
interior to a radius of influence of the satellite galaxy.   We
introduced the free parameter $\psi$ that dictated how the radius of
influence could expand or shrink allowing for more or less stellar
mass loss.  We calculated the values of $\epsilon$ and $\psi$ that
matched the observed clustering of SDSS galaxies over a large range of
luminosity threshold samples.  Matching to the observed clustering at each
luminosity threshold directly informs us of the amount of stellar mass lost
from satellite galaxies and, thus, will comprise the IHL.  

This procedure enables predictions for the amount of IHL or IHL+BHG compared to the total
stellar mass of a host system of a given mass ($\fihl$ and $\fihlbhg$,
respectively).  Observationally, it has been established that the
amount of diffuse material is nearly negligible at the scale of
individual galaxies
\citep[e.g.,][]{sackett94,morrison97,weil97,lequeux98,abe99,zibetti_ferguson04,bailin11}.
For example, M33, which is $\sim 10^{11}\ \Msun$, has been estimated
to have an $\fihl \leq 1\%$ \citep{McConnachie06,hood07}.   Our own
Milky Way is thought to have an $\fihl \sim 1\%$
\citep[e.g.,][]{morrison93,wetterer_mcgraw96,chiba_beers00,ivezic00,morrison00,yanny00,siegel02}
or $\sim 2\%$ when including the stars from the Sagittarius stream
\citep{law05}. Our analog M31 has an $\fihl \sim 2-5\%$
\citep{irwin05,guhathakurta05,chapman06,kilarai06}.  Recently,
\citet{bailin11} showed that NGC 235, which is of comparable
luminosity to the Milky Way and M31, has $\fihl \sim 6\%$.   On group
scales, $\sim 10^{13} - 10^{14}\ \Msun$, measured $\fihl$ percentages
tend to be higher than on galaxy scales, though there can be
considerable variation from group to group (e.g., $\fihl$ on the order
of a few percent: \citealt{feldmeier01,castrorodriguez03,feldmeier06},
$\fihl \sim 5-30\%$: \citealt{darocha05,aguerri06,darocha08}, and as
high as $\fihl \sim 45\%$: \citealt{white_P03,mcgee_balogh10}).  For
clusters scales, IHL fractions are typically much higher.  There is
still substantial scatter, but $\fihl$ values range from $\sim 10 -
40\%$
\citep{melnick77,thuan_kormendy77,uson91,bernstein95,calc_rold00,lin_mohr04,mihos05,zibetti05,krick06,seigar07}.
As discussed in \S \ref{sec:predict_IHL}, \citet{gonzalez07} posit
that the relevant quantity is $\fihlbhg$, because it can be difficult
to disentangle the stellar mass associated with  central galaxy from
the diffuse IHL component.  It has been found that $\fihlbhg \sim
30\%$ on average on cluster scales
\citep{zibetti05,gonzalez07,pierini08,toledo11}.

Our modeling of satellite galaxy stellar mass loss 
has yielded the following principal results and conclusions.
\begin{itemize}
\item
Abundance matching with $\epsilon = 0$ and $\psi = 0$ (``traditional''
abundance matching, akin to using the subhalo mass at the time of
infall) over-predicts the correlation function on small
scales ($\lesssim 1\Mpc$) for each luminosity threshold sample.
\item
For each luminosity threshold, we found both the $\epsilon$ and $\psi$
value that predicted a correlation function that matched the data.  We
found that $\epsilon$ and $\psi$  were decreasing functions of 
luminosity.  This means that low-luminosity satellite galaxies are more 
efficiently stripped of stellar material than luminous satellites.

\item
Our predictions for $\fihl$ from both Model~1 and 
Model~2 are in good agreement with observational
studies over an enormous range of host halo mass scales. For host
masses of $10^{12}$ (individual galaxy scale), $10^{13} - 10^{14}$
(group to small cluster scales), and $10^{14} - 10^{15}$ (cluster
scales), we find mean IHL fractions of a few percent, $10-20\%$, and
$20 - 25\%$, respectively.  These are in accord with previous
observational studies, though there is substantial scatter in the
$\fihl$ measurements in the literature.
\item
The more physically-motivated Model~2 is consistent with the
$\fihlbhg$ measurements of \citet{gonzalez07} from small group scales
all the way through cluster mass systems.  The scatter in the Model~2
results is comparable to the scatter observed in the data.  On the 
contrary, the scatter among $\fihlbhg$ values predicted by Model~1 is 
insufficient to describe the scatter among the observed systems.  
This suggests that Model~1, in which stellar mass loss occurs in 
proportion to dark  matter mass loss, can be rejected by current data.
\end{itemize}

We have shown that \emph{galaxy clustering} can be used as a powerful
tool to understand how satellite galaxies lose stellar mass.  We have
found the interesting result that low-luminosity galaxies lose more
stellar mass relative to subhalo dark matter mass loss than luminous
galaxies.  Moreover, we were able to predict current IHL observations
and thus further constrain our stellar mass loss models.  These
results show that our modeling framework can mitigate potential
problems associated with the conventional abundance matching approach.
Also, our approach is generalizable.  It allows for the flexibility to
make more  detailed predictions within this framework (for instance,
examining  the photometric properties of the IHL as compared to
galaxies).

In Paper~II, we will take advantage of clustering measurements  at
high redshifts, allowing us to study the evolution of satellite galaxy
stellar mass loss as a function of \emph{time}.  This will enable us
to make predictions for the build-up of the IHL over cosmic time.   A
possible extension of this program is to also explore stellar mass-selected
threshold samples directly rather than the luminosity threshold
samples we have used in this paper.  This is a natural choice for
refining this class of studies because dynamical models most directly
treat stellar mass loss.

A further useful avenue to pursue as a result of this work will be to
connect  detailed theoretical models of stellar mass loss in
individual galaxies more directly to the statistical models of stellar
mass loss that can be explored with large-scale survey data, such as
we have done.  Models of stellar mass loss that are consistent with
survey data must also be representative of detailed dynamical models
of stellar mass loss.  It will be very interesting to develop a set of
simple, yet powerful models for the build-up and dispersal of the
stellar mass in satellite galaxies that simultaneously describe the
evolution of galaxy clustering and intrahalo light over cosmic time.


\section{acknowledgments}
We thank Chris Purcell and Jeff Newman for insightful discussions.
The work of DFW and AAB is supported by Vanderbilt University, the
Alfred P. Sloan Foundation, and the National Science Foundation (NSF)
through grant AST-1109789.  The work of ARZ is funded by the
Pittsburgh Particle  physics, Astrophysics, and Cosmology Center (PITT
PACC) at the University of  Pittsburgh and by the NSF through grants
AST-0806367 and  AST-1108802.  


\bibliography{./citations.bib}

\end{document}